\documentclass[preprint,authoryear,12pt]{elsarticle}

\usepackage{amssymb}
\usepackage{graphicx}
\usepackage{epsfig}
\usepackage{array}
\usepackage{mathrsfs}
\usepackage{amsfonts}
\usepackage{amsmath}
\usepackage{amssymb}
\usepackage{mathrsfs}
\input xy
\xyoption{all}

\usepackage{tipa}
\usepackage{pifont}
\usepackage{cases}
\usepackage{txfonts}
\usepackage{graphicx}
\usepackage{array}
\usepackage{mathrsfs}
\usepackage{amsfonts}
\usepackage{setspace}
\usepackage{subfigure}

\newtheorem{theorem}{Theorem}

\newdefinition{definition}{Definition}

\newtheorem{remark}{Remark}
\newtheorem{example}{Example}

\newproof{pf}{Proof}
\newproof{pot}{Proof of Theorem \ref{thm2}}

\journal{Automatica}

\begin{document}

\begin{frontmatter}


 \title{Decentralized Supervisory Control of Discrete Event Systems for Bisimulation Equivalence \tnoteref{label1}}

 \author{Yajuan Sun}
 \ead{sunyajuan@nus.edu.sg}
 \author{Hai Lin}
 \ead{elelh@nus.edu.sg}
 \author{Ben M. Chen}
\ead{bmchen@nus.edu.sg}
\address{Department of Electrical and Computer Engineering, National University of Singapore,
Singapore, 117576}
%



\begin{abstract}
In decentralized systems, branching behaviors naturally arise due
to communication, unmodeled dynamics and system abstraction, which
can not be adequately captured by the traditional sequencing-based
language equivalence. As a finer behavior equivalence than
language equivalence, bisimulation not only allows the full set of
branching behaviors but also explicitly specifies the properties
in terms of temporal logic such as CTL* and mu-calculus. This
observation motivates us to consider the decentralized control of
discrete event systems (DESs) for bisimulation equivalence in this
paper, where the plant and the specification are taken to be
nondeterministic and the supervisor is taken to be deterministic.
An automata-based control framework is formalized, upon which we
develop three architectures with respect to different decision
fusion rules for the decentralized bisimilarity control, named a
conjunctive architecture, a disjunctive architecture and a general
architecture. Under theses three architectures, necessary and
sufficient conditions for the existence of decentralized
bisimilarity supervisors are derived respectively, which extend
the traditional results of supervisory control from language
equivalence to bisimulation equivalence. It is shown that these
conditions can be verified with exponential complexity.
Furthermore, the synthesis of bisimilarity supervisors is
presented when the existence condition holds.
\end{abstract}


\end{frontmatter}


\label{}

\section{INTRODUCTION}

A decentralized system is composed of many distributed and
networked local agents, in which each local agent makes control
decisions based on their own information and then forms a global
decision to induce the system to achieve a desired behavior. Lots
of physical systems, such as communication systems
\citep{cieslak1988supervisory}, \citep*{rudie1990supervisory},
manufacturing systems \citep*{lin1988decentralized} and networked
computer systems \citep{jensen1992asynchronous},
\citep{ferguson1996economic}, are examples of decentralized
systems. Therefore, the decentralized DESs control problem has
been received increasing attentions with growing engineering
demands in recent years.

The decentralized control of discrete event systems was firstly
addressed by \cite{rudie1992think} under a $C\&P$ (conjunctive and
permissive) architecture. While some other works considered the
decentralized supervision problem by using different
architectures. For example, a $D \& A$ (disjunctive and
anti-permissive) architecture was presented in
\citep{yoo2002general}, which is complementary with the
conjunctive architecture. To generalize the $C \&P$ architecture
and the $D \& A$ architecture, \cite{yoo2002general} proposed a
general architecture, which combines above mentioned two
architectures. In \citep{yoo2004decentralized}, a conditional
architecture was used for allowing the controller to take
conditional decisions. In \citep{ricker2000know}, a
knowledge-based architecture was provided to associate the
decision of the supervisor to a grade or level of ambiguity. Based
on these architectures, recent works investigated the hierarchical
control \citep{schmidt2008nonblocking}, the reliable control
\citep{takai2000reliable}, \citep{liu2010reliable} and the
communicating control with \citep{park2007technical} or without
\citep{barrett2000decentralized}, \citep{van2004decentralized}
communication delays for decentralized supervisory control of
DESs. All these work employed language equivalence as the notion
of behavior equivalence. However, the traditional sequencing-based
language equivalence is not adequate for branching behaviors which
naturally arise due to communication, synchronization, unmodeled
dynamics and system abstraction. This calls for the development of
a new decentralized supervisory control framework that can fully
capture the branching information, while at the same time
possesses a practical implementation complexity.

In this paper, we adopt the bisimulation relation as the behavior
equivalence between controlled system and specifications. As a
finer behavior equivalence than language equivalence, bisimulation
was introduced by \cite{milner} and \cite{park1981concurrency},
since then it has been successfully used in model checking
\citep{clarke1997model}, software verification
\citep{chaki2004abstraction} and formal analysis of continuous
\citep{antoniotti2004taming}, \citep{desharnais2002bisimulation},
\citep{kloetzer2007temporal}, \citep{tabuada2004bisimilar}, hybrid
\citep{haghverdi2005bisimulation} and discrete event systems. More
appealing, bisimulation allows the full set of branching behaviors
and explicitly specifies the properties in terms of temporal logic
such as CTL* \citep*{emerson1990temporal} and mu-calculus
\citep*{basu2006quotient} while language equivalence only
preserves the linear temporal logic (LTL)-a subclass of CTL*. This
observation strongly motivates us to consider the decentralized
control of DESs for bisimulation equivalence.

The use of bisimulation for DESs subject to language equivalence
was explored in \citep{barrett1998bisimulation},
\citep{rutten1999coalgebra}, \citep{komenda2005control}, and
\citep{sumodel}. The control of DESs for achieving bisimulation
equivalence was studied by \cite{madhusudan2002branching},
\cite{sun2011bis}, \cite{tabuada2004open},
\cite{tabuada2008controller}, \cite{zhou2007small},
\cite{zhoubisimilarity} and \cite{liu2011bisimilarity}. It is
worthy mentioning that all existing work on bisimularity
supervisory control focused on the centralized control framework.

To the best our knowledge, no prior work considered the
decentralized control of DESs for bisimulation equivalence. The
contributions of this paper mainly lie on the following aspects.
Firstly, a novel automata-based framework is proposed to address
the decentralized bisimilarity supervisory control problem. For
such a framework, all of the plant, the specification and the
supervised system are modeled as automata and allowed to be
nondeterministic. Accordingly, the decentralized bisimilarity
supervisor is formalized by an automaton and a local decision map,
in which the automaton dynamically tracks and synchronizes the
behaviors of the plant and the local decision map determines
whether enables the events defined at the state of the automaton
or not. Based on different local decision maps and global decision
fusion rules, the decentralized bisimilarity control problem can
be developed with three architectures-a conjunctive architecture,
a disjunctive architecture and a general architecture. Secondly,
to effectively implement the proposed strategy, deterministic
supervisors are our focus in this paper. We provide the necessary
and sufficient conditions for the existence of bisimilarity
supervisors for above three architectures respectively, which
extends the traditional results of supervisory control from
language equivalence to bisimulation equivalence. It is shown that
these conditions can be verified with exponential complexity.
Furthermore, the obtained results illustrate that the conjunctive
architecture is complementary with the disjunctive architecture
(See Example 2 and Example 3) and both of them are special cases
of general architecture (see Example 4), which coincides with the
cases for language equivalence. Thirdly, when the existence
condition holds, we also present the methods to design the
decentralized bisimilarity supervisors for the proposed three
architectures.


The rest of paper is organized as follows. Section 2 gives the
preliminary. Section 3 presents the problem formulation. The
decentralized bisimilarity control problem for the conjunctive
architecture, the disjunctive architecture and the generalized
architecture are explored in Section 4, Section 5 and Section 6
respectively. Illustrative examples are provided in Section 7. The
paper concludes with section 8.

\section{Preliminary}
A nondeterministic DES is modeled as an automaton $G
=(X,\Sigma,x_{0},\alpha, X_{m})$, where $X$ is the set of states,
$\Sigma$ is a finite set of events, $\alpha: X \times \Sigma
\rightarrow 2^X$ is the transition function, $x_0$ is the initial
state, and $X_m \subseteq X$ is the set of marked states. The
active event set at state $x$ is defined as $E_{G}(x)=\{\sigma \in
\Sigma~|~\alpha(x, \sigma)$ is defined\}. Let $\Sigma^{*}$ be the
set of all finite strings over $\Sigma$, including the empty
string $\epsilon$. The transition function $\alpha$ can be
extended to $\alpha: X \times \Sigma^{*} \rightarrow 2^{X}$ in the
natural way: $\alpha(x, \epsilon)=x$; $\alpha(x,
s\sigma)=\alpha(\alpha(x, s), \sigma)$ for $s\in \Sigma^{*}$ and
$\sigma \in \Sigma$. If the transition function is a partial map
$\alpha: X \times \Sigma \rightarrow X$, the DES is said to be
deterministic. Given a string $s \in \Sigma^{*}$, $|s|$ is the
length of the string $s$. The language generated by $G$ is defined
as $L(G)=\{s \in \Sigma^{*} \mid \alpha(x_0, s)$ is defined$\}$,
and the marked language is defined as $L_{m}(G)=\{s \in \Sigma^{*}
\mid \alpha(x_0, s) \cap X_m \neq \emptyset$\}. The event set can
be partitioned into $\Sigma$ = $\Sigma_{uc}\dot{ \cup}
\Sigma_{c}$, where $\Sigma_{uc}$ is the set of uncontrollable
events and $\Sigma_{c}$ is the set of controllable events. Under
partial observation, it can be also partitioned into $\Sigma$ =
$\Sigma_{uo} \dot{\cup} \Sigma_{o}$, where $\Sigma_{uo}$ is the
set of unobservable events and $\Sigma_{o}$ is the set of
observable events. When a string of events occurs, the sequence of
observed events is filtered by a projection $P$: $\Sigma^{*}
\rightarrow \Sigma_{o}^{*}$, which is defined inductively as
follows: $P(\epsilon)=\epsilon$, for $\sigma \in \Sigma$ and $s
\in \Sigma^{*}$, $P(s\sigma)=P(s)\sigma$ if $\sigma \in \Sigma_o$,
otherwise, $P(s\sigma)=P(s)$. Consider a language $K$. The prefix
closure of $K$, denoted as $\overline{K}$, is the language
$\overline{K}=\{s \in \Sigma^{*}~|~(\exists t \in
\Sigma^{*})~st\in K\}$. The Kleene closure of $K$, denoted as
$K^{*}$, is the language $K^*=\cup_{n \in \mathbb{N}}K^{n}$, where
$K^{0}=\epsilon$ and for each $n \geq 0$, $K^{n+1}=K^{n}K$. For a
nondeterministic $G$, let $det(G)$ be a minimal deterministic
automaton such that $L(det(G))=L(G)$ and $L_{m}(det(G))=L_{m}(G)$.




To synchronize the automata, the product operator is introduced as
below \citep*{cassandras2008introduction}.

\begin{definition}
Given $G_1 =(X_1,\Sigma,x_{01},\alpha_1, X_{m1})$ and $G_2
=(X_2,\Sigma,x_{02},\alpha_2,X_{m2})$, the product of $G_1$ and
$G_2$ is an automaton
\[
G_1 || G_2 = ( X_1 \times X_2, \Sigma, \alpha_{1||2}, (x_{01},
x_{02}), X_{m1} \times X_{m2}),
\]
where for $x_1 \in X_1$, $x_2 \in X_2$ and $\sigma \in \Sigma$,
the transition function is defined as:

\[
\alpha_{1||2}((x_1, x_2),\sigma) = \left\{ {\begin{array}{*{20}c}
   \alpha_1(x_1, \sigma) \times \alpha_2(x_2, \sigma) & {\sigma \in E_{G_1}(x_1) \cap E_{G_2}(x_2) };  \\
    \emptyset & {otherwise}.  \\
\end{array}} \right.
\]
\end{definition}

In the conventional supervisory control problem, language
controllability \citep*{ramadge1984supervisory} is a necessary and
sufficient condition for the existence of a supervisor that
achieves language equivalence between the supervised system and
the specification, and it is captured by the following definition.

\begin{definition}\label{langc}
Consider an automaton $G =(X,\Sigma,x_{0},\alpha, X_{m})$, where
$\Sigma_{uc} \subseteq \Sigma$ is the set of uncontrollable
events. A language $K \subseteq L(G)$ is said to be language
controllable with respect to $L(G)$ and $\Sigma_{uc}$ if
\[
\overline{K}\Sigma_{uc} \cap L(G) \subseteq \overline{K}.
\]
\end{definition}

Bisimulation is a finer behavior equivalence than language
equivalence, which is stated as follows \citep*{milner}. It is
well known that bisimulation equivalence implies language
equivalence and marked language equivalence, but the converse does
not hold.


\begin{definition}
Given $G_{1} =(X_{1},\Sigma,x_{01},\alpha_{1},X_{m1})$ and $G_{2}
=(X_{2},\Sigma,x_{02},\alpha_{2},X_{m2})$, a simulation relation
$\phi$ is a binary relation $\phi \subseteq X_1 \times X_2$ such
that $(x_{1}, x_{2}) \in \phi$ implies
\begin{enumerate}
\item[(1)] $(\forall \sigma
\in \Sigma)[\forall x_{1}^{'} \in
\alpha_{1}(x_{1},\sigma)\Rightarrow \exists x_{2}^{'} \in
\alpha_{2}(x_{2},\sigma)$ such that $(x_{1}^{'},x_{2}^{'}) \in
\phi]$;

\item[(2)] $x_{1} \in X_{m1} \Rightarrow x_{2} \in X_{m2}$.
\end{enumerate}
\end{definition}

The automaton $G_{1}$ is said to be simulated by $G_{2}$, denoted
by $G_{1} \prec_{\phi} G_{2}$, if there is a binary relation
$\phi$ $\subseteq$ $X_{1} \times X_{2}$ such that $(x_{01},x_{02})
\in \phi$. For $\phi \subseteq (X_1 \cup X_2) \times (X_1 \cup
X_2)$, if $G_{1} \prec_{\phi} G_{2}$, $G_{2} \prec_{\phi} G_{1}$
and $\phi$ is symmetric, $\phi$ is called a bisimulation relation
between $G_{1}$ and $G_{2}$, denoted by $G_{1} \cong_{\phi}
G_{2}$. We sometimes omit the subscript $\phi$ from $\prec_{\phi}$
or $\cong_{\phi}$ when it is clear from the context.


\section{Problem Formulation}



A nondeterministic system $G$ is jointly controlled by $n$ local
supervisors $\mathcal{S}_{1}$, $\mathcal{S}_{2}$ $\cdots$
$\mathcal{S}_{n}$ for achieving the bisimulation equivalence
between the supervised system and the given nondeterministic
specification $R$. A priori information available to each local
supervisor includes the desired behavior $R$ and the decision
fusion rule to form a global decision. Further, each local
supervisor can observe the locally observable information and
control the locally controllable events.

Denote $\Sigma_{ci}$ and $\Sigma_{uci}$ as the sets of locally
controllable and uncontrollable events respectively; $\Sigma_{oi}$
and $\Sigma_{uoi}$ as the sets of locally observable and
unobservable events, respectively, where $i \in I=\{1, 2, \cdots,
n\}$. The sets of globally controllable and globally observable
events are defined as $\Sigma_{c}=\cup_{i \in I}\Sigma_{ci}$ and
$\Sigma_{o}=\cup_{i \in I}\Sigma_{oi}$ respectively. Then,
$\Sigma_{uc}=\Sigma -\Sigma_{c}$ is the set of globally
uncontrollable events and $\Sigma_{uo}=\Sigma -\Sigma_{o}$ is the
set of globally unobservable events.

The local supervisor $\mathcal{S}_{i}$ is defined as a tuple
\begin{equation}\label{sup}
\mathcal{S}_{i}=(S_i, \psi_{i}),
\end{equation}

where $S_{i}=(Y_{i}, \Sigma, \beta_{i}, y_{0i}, Y_{mi})$ is an
automaton with $Y_{mi}=Y_{i}$ and $\psi_{i}: Y_{i} \rightarrow
2^{\Sigma}$ is the local decision map.

It can be seen that a local supervisor consists an automaton
$S_{i}$ and a local decision map $\psi_{i}$, in which $S_{i}$
dynamically tracks and synchronizes the behaviors of the plant and
$\psi_{i}$ determines whether enables the events defined at the
state of $S_{i}$ or not. A local supervisor is called to be
nondeterministic if $S_i$ is nondeterministic, otherwise, it is
called to be deterministic. To reduce the implementation
complexity, local supervisors adopted in this paper are assumed to
be deterministic.

Because a local supervisor possesses limit control and observation
capabilities, an admissible local supervisor should satisfy the
following properties.




\begin{definition}
Consider a local supervisor $\mathcal{S}_{i}=((Y_{i}, \Sigma,
\beta_{i}, y_{0i}, Y_{mi}), \psi_{i})$. Then,

\begin{itemize}
\item $\mathcal{S}_{i}$ is called $\Sigma_{uoi}-compatible$ if
$\forall y \in Y_{i}$ and $\forall \sigma \in \Sigma_{uoi}$,
$\beta_{i}(y, \sigma)=y$;

\item $\mathcal{S}_{i}$ is called $\Sigma_{uci}-compatible$ if
$\forall y \in Y_{i}$ and $\forall \sigma \in \Sigma_{uci}$,
$\beta_{i}(y, \sigma)\neq \emptyset$;

\item $\mathcal{S}_{i}$ is called $(\Sigma_{uoi},
\Sigma_{uci})-compatible$ if it is $\Sigma_{uoi}-compatible$ and
$\Sigma_{uci}-compatible$.



\end{itemize}
\end{definition}

That is, a $\Sigma_{uoi}-compatible$ local supervisor does the
same control actions for the indistinguishable events and a
$\Sigma_{uci}-compatible$ supervisor defines local uncontrollable
events at each state of the automaton.

Further, the decisions from local supervisors can be synthesized
through the decision fusion rule, which is stated as follows.

\begin{definition}
Given local supervisors $\mathcal{S}_{i}=(S_i, \psi_{i})$ with
$||_{i\in I}S_i=(Y_{||}, \Sigma, \beta_{||}, y_{0||}, Y_{m||})$,
where $i \in I$, the decision fusion rule $\psi_{f}$ is defined as
\begin{equation}\label{glocal dec}
\psi_{f}: Y_{||} \rightarrow \Gamma:=\{\gamma \in 2^{\Sigma}:
\Sigma_{uc} \subseteq \gamma\}.
\end{equation}
\end{definition}

Then, the decentralized bisimilarity control of discrete event
systems can be classified with respect to different decision
fusion rules.

In the rest of this paper, we will use $G=(X, \Sigma, \alpha,
x_{0}, X_m)$, $R=(Q, \Sigma, \delta, q_{0}, Q_m)$,
$\mathcal{S}_{i}=(S_i, \psi_{i})=((Y_{i}, \Sigma, \beta_{i},
y_{0i}, Y_{mi}), \psi_{i})$ and $||_{i \in I}S_i=(Y_{||}, \Sigma,
\beta_{||}, y_{0||}, Y_{m||})$ to denote the nondeterministic
plant, the nondeterministic specification, the local supervisor
and the product of $S_{i}$ respectively unless otherwise stated.



\subsection{Conjunctive Architecture}
In the subsection, a conjunctive decision fusion approach is
presented for synthesizing decisions of local supervisors.

For a conjunctive architecture, a local supervisor
$\mathcal{S}_{i}$ enables $\Sigma \setminus \Sigma_{ci}$ by
default, i.e., $\Sigma \setminus \Sigma_{ci} \subseteq
\psi_{i}(y)$ for any $y \in Y_{i}$. Then, the conjunctive decision
fusion rule is expressed as follows.



\begin{definition}
Given local supervisors $\mathcal{S}_{i}=(S_i, \psi_{i})$ with
$||_{i\in I}S_i=(Y_{||}, \Sigma, \beta_{||}, y_{0||}, Y_{m||})$,
where $i \in I$, the conjunctive decision fusion rule $\psi_{fc}:
Y_{||} \rightarrow 2^{\Sigma}$ is defined as
\begin{equation}\label{conj global dec}
\psi_{fc}(y_1, y_2, \cdots, y_{n})=\cap_{i \in I} \psi_{i}(y_{i}).
\end{equation}
\end{definition}

In the conjunctive architecture, $C \&P$ co-observability is the
key property for the existence of a set of local supervisors to
control the plant to be language equivalent to the specification
\citep{rudie1992think}.

\begin{definition}
Given a plant $G$, a language $K \subseteq L(G)$ is said to be $C
\&P$ co-observable with respect to $L(G)$, $\Sigma_{oi}$ and
$\Sigma_{ci}$, where $i \in I$, if for any $s \in \overline{K}$
and $\sigma \in \Sigma_{c}$ such that $s\sigma \in L(G) \setminus
\overline{K}$,
\begin{equation}
(\exists i \in I)[(P_{i}^{-1}P_{i}(s)\sigma \cap \overline{K} \neq
\emptyset)\wedge (\sigma \in \Sigma_{ci})].
\end{equation}
\end{definition}


\subsection{Disjunctive Architecture}
This subsection introduces a disjunctive architecture, in which a
disjunctive decision fusion rule is employed for the supervisor
decision synthesis.

For such a disjunctive architecture, a local supervisor
$\mathcal{S}_{i}$ disables $\Sigma \setminus \Sigma_{ci}$ by
default, i.e., $\Sigma \setminus \Sigma_{ci} \cap \psi_{i}(y)
=\emptyset$ for any $y \in Y_{i}$. And the disjunctive fusion rule
is presented as below.


\begin{definition}
Given local supervisors $\mathcal{S}_{i}=(S_i, \psi_{i})$ with
$||_{i\in I}S_i=(Y_{||}, \Sigma, \beta_{||}, y_{0||}, Y_{m||})$,
where $i \in I$, the disjunctive decision fusion rule $\psi_{fd}:
Y_{||} \rightarrow 2^{\Sigma}$ is defined as

\begin{equation}\label{disj global dec}
\psi_{fd}(y_1, y_2, \cdots, y_{n})=\cup_{i \in I} \psi_{i}(y_{i}).
\end{equation}
\end{definition}

Then, we state the notion of $D \&A$ co-observability
\citep{yoo2002general}, which guarantees the existence of
decentralized language equivalence enforcing supervisors with a
disjunctive structure.

\begin{definition}
Given a plant $G$, a language $K \subseteq L(G)$ is said to be $D
\&A$ co-observable with respect to $L(G)$, $\Sigma_{oi}$ and
$\Sigma_{ci}$, where $i \in I$, if for any $s \in \overline{K}$
and $\sigma \in \Sigma_{c}$ such that $s\sigma \in \overline{K}$,
\begin{equation}
(\exists i \in I)[((P_{i}^{-1}P_{i}(s)\cap \overline{K})\sigma
\cap L(G) \subseteq \overline{K})\wedge (\sigma \in \Sigma_{ci})].
\end{equation}
\end{definition}


\subsection{General Architecture}
In the general architecture, the event set $\Sigma_{c}$ is further
partitioned into $\Sigma_{c}=\Sigma_{ce} \cup \Sigma_{cd}$, where
$\Sigma_{ce}$ is the set of controllable events which is enabled
by default in local decision and $\Sigma_{cd}$ is the set of
controllable events which is disabled by default in local
decision. That is, a local supervisor $\mathcal{S}_{i}$ for a
general architecture satisfies $\Sigma_{ce} \setminus \Sigma_{ci}
\subseteq \psi_{i}(y)$ and $\Sigma_{cd} \setminus \Sigma_{ci} \cap
\psi_{i}(y) =\emptyset$ for any $y \in Y_i$. Denote
$\Sigma_{cei}=\Sigma_{ci} \cap \Sigma_{ce}$ and
$\Sigma_{cdi}=\Sigma_{ci} \cap \Sigma_{cd}$.


Further, the decision fusion rule of the general architecture is
captured by the following definition.

\begin{definition}
Consider local supervisors $\mathcal{S}_{i}=(S_i, \psi_{i})$ with
$||_{i\in I}S_i=(Y_{||}, \Sigma, \beta_{||}, y_{0||}, Y_{m||})$,
where $i \in I$. The generalized decision fusion rule $\psi_{fg}:
Y_{||} \rightarrow 2^{\Sigma}$ is defined as
\begin{equation}\label{general global dec}
\psi_{fg}(y_1, y_2, \cdots, y_{n})= P_{\Sigma_{ce}}[\cap_{i \in I}
\psi_{i}(y_{i})] \cup P_{\Sigma_{cd}}[\cup_{i \in I}
\psi_{i}(y_{i})] \cup \Sigma_{uc},
\end{equation}
where $P_{\Sigma_{ce}}: \Sigma^{*} \rightarrow \Sigma_{ce}^{*}$
and $P_{\Sigma_{cd}}: \Sigma^{*} \rightarrow \Sigma_{cd}^{*}$ are
projections.
\end{definition}



With this general architecture, the following concept is used as
the existence condition for decentralized supervisors to achieve
the language equivalence between the plant and the specification
\citep{yoo2002general}.

\begin{definition}
Given a plant $G$, a language $K \subseteq L(G)$ is said to be
co-observable with respect to $L(G)$, $\Sigma_{oi}$ $\Sigma_{cei}$
and $\Sigma_{cdi}$, where $i \in I$ if

\begin{enumerate}

\item[(1)] $K$ is $C\&P$ co-observable with respect to $L(G)$,
$\Sigma_{oi}$ and $\Sigma_{cei}$ for $i\in I$;

\item[(2)] $K$ is $D\&A$ co-observable with respect to $L(G)$,
$\Sigma_{oi}$ and $\Sigma_{cdi}$ for $i\in I$.
\end{enumerate}
\end{definition}


Next, the decentralized supervised system for bisimulation
equivalence is introduced as below.

\begin{definition}\label{supvised system}
Consider a plant $G=(X, \Sigma, \alpha, x_0, X_{m})$, a
specification $R = (Q, \Sigma, \delta, q_0, Q_{m})$, local
supervisors $\mathcal{S}_{i}=(S_i, \psi_{i})$ with $||_{i \in
I}S_{i}=(Y_{||}, \Sigma, \beta_{||}, y_{0||}, Y_{m||})$ and a
decision fusion rule $\psi_{f}$, where $i \in I$. The supervised
system $cl_{i\in I}(S_{i}, \psi_{i})/_{\psi_{f}}G$ is defined as
an automaton
\begin{equation}
cl_{i\in I}(S_{i}, \psi_{i})/_{\psi_{f}}G = (X', \Sigma, \alpha',
x_0', X_{m}'),
\end{equation}
where $X' \subseteq X \times Y_{||}$ is the set of states
reachable from the initial state $x_0'=(x_0, y_{0||})$, $X_{m}'
\subseteq X_{m} \cap Y_{m||}$ and the transition function
$\alpha': X' \times \Sigma^{*} \rightarrow 2^{X'}$ is defied
inductively:

\begin{enumerate}
\item[(1)] $(x, y) \in \alpha'(x_0', \sigma) \Leftrightarrow x \in
\alpha(x_0, \sigma) \wedge y \in \beta_{||}(y_{0||}, \sigma)
\wedge \sigma \in \psi_{f}(y_{0||})$;

\item[(2)] If $(x, y) \in \alpha'(x_0', s)$, then $(x', y') \in
\alpha'((x, y), \sigma) \Leftrightarrow  x' \in \alpha(x, \sigma)
\wedge y' \in \beta_{||}(y, \sigma) \wedge \sigma \in
\psi_{f}(y)$.
\end{enumerate}
\end{definition}

The decision fusion rule $\psi_{f}$ can be in terms of $\psi_{fc}$
(\ref{conj global dec}), $\psi_{fd}$ (\ref{disj global dec}) or
$\psi_{fg}$ (\ref{general global dec}). Therefore, the supervised
system can be adopted for all of the conjunctive architecture, the
disjunctive architecture and the general architecture. Moreover,
this supervision framework can be easily implemented as below.
When a certain event occurs in the plant, the automata of local
supervisors will update to new states based on their own
observation. At these states, local decisions are made and then
fuse a global decision which will be delivered to the plant
through a communication channel to enforce a desired behavior.

\begin{remark}
With respect to language equivalence, the supervised system is
defined inductively based on strings in \citep{rudie1992think} and
\citep{yoo2002general}. In this paper, bisimulation equivalence is
our focus, and it allows the nondeterminism of the plant, the
specification and the supervisor. Thus, we generalizes the
string-based description to the automata-based description.
\end{remark}

Based on the proposed frameworks, this paper aims to tackle the
following decentralized bisimilarity control problem:

\emph{Given a plant $G$ and a specification $R$ modeled as
nondeterministic discrete event systems with $L(R) \subseteq
L(G)$, does there exist a set of $(\Sigma_{uoi},
\Sigma_{uci})-compatible$ supervisors $\mathcal{S}_{i}=(S_{i},
\psi_{i})$ where $i \in I$ such that $cl_{i\in I}(S_{i},
\psi_{i})/_{\psi_{f}}G \cong R$ for the conjunctive architecture
($\psi_{f}=\psi_{fc}$), the disjunctive architecture
($\psi_{f}=\psi_{fd}$) and the general architecture
($\psi_{f}=\psi_{fg}$) respectively? If so, how to construct
$\mathcal{S}_{i}$?} 

\section{Conjunctive Architecture}
The decentralized bisimilarity control problem under the
conjunctive architecture is investigated in this section. From
(\ref{sup}) and Definition \ref{supvised system} of the previous
section, we can see that the marking only depends on the plant
because the decentralized bisimilarity supervisor plays no role in
the marking. Thus, the following concept is introduced to
guarantee the existence of decentralized bisimilarity supervisors.

\begin{definition}
Given a plant $G=(X, \Sigma, \alpha, x_0, X_{m})$ and a
specification $R = (Q, \Sigma, \delta, q_0, Q_{m})$ with $L(R)
\subseteq L(G)$, $R$ is called to be marked language closed with
respect to $G$ if
\begin{equation}
(\forall s \in L(R))[s \in L_{m}(G) \Rightarrow s \in L_{m}(R)].
\end{equation}
\end{definition}

Then, the following theorem presents a necessary and sufficient
condition for the existence of $(\Sigma_{uoi},
\Sigma_{uci})-compatible$ bisimilarity supervisors under the
conjunctive architecture.


\begin{theorem}\label{conjunctive}
Given a plant $G=(X, \Sigma, \alpha, x_0, X_{m})$ and a
specification $R = (Q, \Sigma, \delta, q_0, Q_{m})$ with $L(R)
\subseteq L(G)$, there exist $(\Sigma_{uoi},
\Sigma_{uci})-compatible$ supervisors $\mathcal{S}_{i}=(S_{i},
\psi_{i})$ with the conjunctive decision fusion rule $\psi_{fc}$
such that $cl_{i\in I}(S_{i}, \psi_{i})/_{\psi_{fc}}G \cong R$ if
and only if the following conditions hold:

\begin{enumerate}
\item[(1)] There is a bisimulation relation $\phi$ such that $G ||
det(R) \cong_{\phi} R$;

\item[(2)] $L(R)$ is language controllable with respect to $L(G)$
and $\Sigma_{uc}$;

\item[(3)] $L(R)$ is $C\&P$ co-observable with respect to $L(G)$,
$\Sigma_{ci}$ and $\Sigma_{oi}$, where $i\in I$;

\item[(4)] $R$ is marked language closed with respect to $G$.
\end{enumerate}
\end{theorem}

\begin{pf}
Consider $det(R)=(Z, \Sigma, \delta_{Z}, \{q_0\}, Z_{m})$ and
$G||det(R)=(X_{XZ}, \Sigma, \alpha_{XZ}, (x_0, \{q_0\}),
X_{mXZ})$.

(Necessity) Let $\mathcal{S}_{i}=(S_{i}, \psi_{i})$=($(Y_{i},
\Sigma, \beta_{i}, y_{0i}, Y_{mi})$, $\psi_{i}$ ) and $||_{i \in
I}S_i=(Y_{||}, \Sigma, \beta_{||}, y_{0||}, Y_{m||})$, where $i
\in I$. Because there is a bisimulation relation $\phi'$ such that
$cl_{i\in I}(S_{i}, \psi_{i})/_{\psi_{fc}}G \cong_{\phi'} R$. We
have $L(cl_{i\in I}(S_{i}, \psi_{i})/_{\psi_{fc}}G) =L(R)$ and
$L_{m}(cl_{i\in I}(S_{i}, \psi_{i})/_{\psi_{fc}}G)=L_{m}(R)$.


We firstly prove that $L(R)$ is language controllable with respect
to $L(G)$ and $\Sigma_{uc}$. For any $s \in L(R)$ and $\sigma \in
\Sigma_{uc}$ such that $s\sigma \in L(G)$, there is $x \in
\alpha(x_0, s)$ with $x' \in \alpha(x, \sigma)$. Because $s \in
L(R)=L(cl_{i\in I}(S_{i}, \psi_{i})/_{\psi_{fc}}G)$ and $||_{i \in
I}S_{i}$ is deterministic, there exists $(x, (y_1, y_2, \cdots,
y_n)) \in \alpha'(x_{0}', s)$. Because $\sigma \in \Sigma_{uc}$,
we have $\sigma \in \psi_{fc}(y_1, y_2, \cdots, y_n)=\cap_{i \in
I}\psi_{i}(y_{i})$. Moreover, $\mathcal{S}_{i}$ is
$\Sigma_{uci}-compatible$, which implies $\beta_{i}(y_i, \sigma)
\neq \emptyset$ for $i \in I$. Thus, there is $(y_1', y_2',
\cdots, y_n') \in \beta_{||}((y_1, y_2, \cdots, y_n), \sigma)$
such that $(x', (y_1', y_2', \cdots, y_n')) \in \alpha'((x, (y_1,
y_2, \cdots, y_n)), \sigma)$ according to Definition \ref{supvised
system}. Therefore, $s\sigma \in L(cl_{i\in I}(S_{i},
\psi_{i})/_{\psi_{fc}}G)=L(R)$.

Secondly, we check $C\&P$ co-observability of $L(R)$ with respect
to $L(G)$, $\Sigma_{ci}$ and $\Sigma_{oi}$, where $i \in I$.
Assume that there is $s \in L(R)$ and $\sigma \in \Sigma_{c}$
satisfying $s\sigma \in L(G)\backslash L(R)$, moreover, either
$\sigma \notin \Sigma_{ci}$ or $P_i^{-1}P_i(s)\sigma \cap L(R)
\neq \emptyset$ for any $i \in I$. For any $j \in I$ satisfies
$\sigma \in \Sigma_{cj}$ and $P_j^{-1}P_j(s)\sigma \cap L(R) \neq
\emptyset$, there exists $s' \in L(R)$ such that
$P_{j}(s)=P_{j}(s')$ and $s'\sigma \in L(R)$. Because
$L(R)=L(cl_{i\in I}(S_{i}, \psi_{i})/_{\psi_{fc}}G)$, we have
$s'\sigma \in L(cl_{i\in I}(S_{i}, \psi_{i})/_{\psi_{fc}}G)$.
Then, there exists $(x, (y_1, y_2 \cdots y_n)) \in \alpha'(x_{0}',
s')$ such that $(x', y') \in \alpha'((x, (y_1, y_2 \cdots y_n)),
\sigma)$. By Definition \ref{supvised system} and $(\ref{conj
global dec})$, we have $\sigma \in \psi_{i}(y_i)$ for $i \in I$.
Since $s \in L(R)$, we have $s \in L(cl_{i\in I}(S_{i},
\psi_{i})/_{\psi_{fc}}G)$. In addition, $s\sigma \in L(G)$. Hence,
there is $x'' \in \alpha(x_0, s)$ such that $(x'', (y_1'', y_2'',
\cdots, y_n'')) \in \alpha'(x_{0}', s)$ and $x''' \in \alpha(x'',
\sigma)$. Because $||_{i \in I}S_{i}$ is deterministic and
$\mathcal{S}_{i}$ is $\Sigma_{uoi}-compatible$, if
$P_{i}(s)=P_{i}(s')$, we have $\beta_{i}(y_{oi},
s)=\beta_{i}(y_{oi}, s')$, where $P_{i}:\Sigma^{*} \rightarrow
\Sigma_{oi}^{*}$ is the projection. For $i \in I$, either $\sigma
\notin \Sigma_{ci}$ or $y_i=y_i''$. Furthermore, $\mathcal{S}_{i}$
is $\Sigma_{uci}-compatible$. Then, there is $y''' \in
\beta_{||}((y_1'', y_2'', \cdots, y_n''), \sigma)$ such that
$(x''', y''') \in \alpha'((x'', (y_1'', y_2'', \cdots, y_n'')),
\sigma)$. It implies $s\sigma \in L(cl_{i\in I}(S_{i},
\psi_{i})/_{\psi_{fc}}G)=L(R)$, which contradicts that $s\sigma
\notin L(R)$. Therefore, the assumption is not correct. Hence,
$L(R)$ is $C\&P$ co-observable with respect to $L(G)$ and
$\Sigma_{ci}$ and $\Sigma_{oi}$.

Thirdly, we verify that there is a bisimulation relation $\phi$
such that $G || det(R) \cong_{\phi} R$. From the definition of
product, we have $L(G||det(R))=L(G) \cap L(det(R))=L(R)$. Thus,
$L(cl_{i\in I}(S_{i}, \psi_{i})/_{\psi_{fc}}G)=L(R)=L(G||det(R))$.
Let $\phi_1=\{((x, z), q) \in X_{XZ} \times Q~|~\exists s\in L(R)$
s.t. $(x, z) \in \alpha_{XZ}((x_0, \{q_0\}), s)$, $q \in
\delta(q_0, s)$, $y \in \beta_{||}(y_{0||}, s)$ and $((x, y), q)
\in \phi'$\}. For any $((x, z), q) \in \phi_1$, if there is a
$\sigma$-successor $(x', z') \in \alpha_{XZ}((x, z), \sigma)$,
where $\sigma \in \Sigma$, we obtain $s\sigma \in L(R)=L(cl_{i\in
I}(S_{i}, \psi_{i})/_{\psi_{fc}}G)$ and $x' \in \alpha(x,
\sigma)$. Because of the determinism of $||_{i\in I}S_{i}$, there
is $y \in \beta_{||}(y_{0||}, s)$ such that $y' \in \beta_{||} (y,
\sigma)$. It implies $(x', y') \in \alpha'((x, y), \sigma)$. Then,
there exists $q' \in \delta(q, \sigma)$ such that $((x', y'), q')
\in \phi'$. Hence, $((x', z'), q') \in \phi_{1}$. If $(x, z) \in
X_{mXZ}$, then $x \in X_{m}$, which implies $(x, y) \in X_{m}'$.
Therefore, $q \in Q_{m}$. For any $(q, (x, z)) \in \phi_1^{-1}$,
if there is a $\sigma$-successor $q'\in \delta(q, \sigma)$, where
$\sigma \in \Sigma$, we have $(x', y') \in \alpha'((x, y),
\sigma)$ such that $((x', y'), q') \in \phi'$ because $((x, y), q)
\in \phi'$. Thus, $x' \in \alpha(x, \sigma)$. Further, $s\sigma
\in L(R)$ implies that there exists $z' \in \delta_{Z}(z, \sigma)$
by the definition of $det(R)$. Thus, $(x', z') \in \alpha_{XZ}((x,
z), \sigma)$. Hence, $(q', (x', z')) \in \phi_{1}^{-1}$. If $q \in
Q_{m}$, then $z \in Z_{m}$ and $x\in X_{m}$. Therefore, $(x, z)
\in X_{mXZ}$. As a result, $G||det(R) \cong_{\phi_1 \cup
\phi_1^{-1}} R$.

Fourthly, we would like to prove that $R$ is marked language
closed with respect to $G$. For any $s\in L(R)$, we have $s \in
L(cl_{i\in I}(S_{i}, \psi_{i})_{\psi_{fc}}/G)$. If $s \in
L_{m}(G)$, there is $x \in X_{m}$ such that $x \in \alpha(x_0,
s)$. Since $s \in L(cl_{i\in I}(S_{i}, \psi_{i})_{\psi_{fc}}/G)$,
we obtain $s \in L_{m}(cl_{i\in I}(S_{i},
\psi_{i})_{\psi_{fc}}/G)$, which implies $s \in L_{m}(R)$.



(Sufficiency) Let $[s]_{i}:=\{s'|P_{i}(s)=P_{i}(s')\}$ for the
projection $P_{i}: \Sigma^{*} \rightarrow \Sigma_{oi}^{*}$. For $i
\in I=\{1, 2, \cdots, n\}$, $\mathcal{S}_{i}=((Y_{i}, \Sigma,
\beta_{i}, y_{0i}, Y_{mi}), \psi_{i})$ is designed as follows:
$Y_{i}=\{z_{di}\} \cup \{[s]_{i}~|~s \in L(R)\}$,
$y_{0i}=[\epsilon]_{i}$, $Y_{mi}=Y_{i}$ and for any $y_{i} \in
Y_{i}$ and $\sigma \in \Sigma$, the transition function
$\beta_{i}$ is defined as:

\begin{equation} \label{con sup}
\beta_{i}(y_{i}, \sigma) = \left\{ {\begin{array}{*{20}c}
  ([s\sigma]_{i}) & {y_{i}=[s]_{i} \wedge
\sigma \in \Sigma_{oi} \wedge [s]_{i}\sigma \cap L(R) \neq \emptyset };  \\ 
 ([s]_{i}) & {y_{i}=[s]_{i} \wedge
\sigma \in \Sigma_{uoi} };  \\
( z_{di}) & {(y_{i}=[s]_{i} \wedge \sigma \in
\Sigma_{uci}\setminus
\Sigma_{uoi} \wedge [s]_{i}\sigma \cap L(R) = \emptyset ) }  \\
& {\vee (y_i=z_{di} \wedge \sigma \in \Sigma_{uoi} \cup
\Sigma_{uci})};\\
    (undefined) & {otherwise}.  \\
\end{array}} \right.
\end{equation}


Further, for any $y_{i} \in Y_{i}$, the local decision map
$\psi_{i}(y_i)$ is defined as:

\begin{equation}\label{con sup dec}
\psi_{i}(y_i) = \left\{ {\begin{array}{*{20}c}
 (\Sigma_{c}\setminus \Sigma_{ci} \cup \Sigma_{uc}
\cup \{\sigma \in \Sigma_{ci}~|~\exists s' \in [s]_{i}, s'\sigma
\in L(R)\}) & {y_{i}=[s]_{i} };  \\ 
(\Sigma_{c}\setminus \Sigma_{ci} \cup \Sigma_{uc}) & {y_i = z_{di} }.  \\
\end{array}} \right.
\end{equation}


Therefore, $\mathcal{S}_{i}$ is $(\Sigma_{uoi},
\Sigma_{uci})-compatible$ and $\psi_{i}$ satisfies the requirement
for the conjunctive architecture. Let $\psi_{fc}$ (\ref{conj
global dec}) be the conjunctive decision fusion rule.

Firstly, we would like to prove that $s \in L(R)$ for any $s \in
L(cl_{i\in I}(S_{i}, \psi_{i})/_{\psi_{fc}}G)$ by the induction
method. (1) $|s|=0$, that is, $s=\epsilon$. We have $\epsilon \in
L(R)$. (2) Suppose that $s \in L(R)$ for any $s \in L(cl_{i\in
I}(S_{i}, \psi_{i})_{\psi_{fc}}/G)$ when $|s|=n$. (3) $|s|=n+1$
with $s=s_1\sigma$. Assume that $s_1\sigma \notin L(R)$. Since
$s_1\sigma \in L(cl_{i\in I}(S_{i}, \psi_{i})_{\psi_{fc}}/G)$,
there is $(x, (y_1, y_2, \cdots, y_n)) \in \alpha'(x_0', s_1)$ and
$\sigma \in \Sigma$ such that $(x', (y_1', y_2', \cdots, y_n'))
\in \alpha'((x, (y_1, y_2, \cdots, y_n)), \sigma)$. Then
$s_1\sigma \in L(G)$ and $\sigma \in \psi_{fc}(y_1, y_2, \cdots,
y_n)=\cap_{i\in I}\psi_{i}(y_i)$. We have the following cases.
Case 1: $\sigma \in \Sigma_{uc}$. Because $|s_1|=n$, we have $s_1
\in L(R)$. Then, $s_1\sigma \in L(R)$ since $L(R)$ is language
controllable with respect to $L(G)$ and $\Sigma_{uc}$. Thus, there
is a contradiction. Case 2: $\sigma \in \Sigma_{c}$. Since $s_1
\in L(R)$, we obtain $y_i=[s_1]_{i}$ for $i \in I$ by the
definition of $\beta_{i}$ and $s_1 \in L(R)$. According to
(\ref{con sup dec}), either $\sigma \notin \Sigma_{ci}$ or there
is $s_1' \in [s_1]_{i}$ such that $s_1'\sigma \in L(R)$, which
violated the $C\&P$ co-observability of $L(R)$ with respect to
$L(G)$, $\Sigma_{oi}$ and $\Sigma_{ci}$, where $i \in I$.
Therefore, the assumption is not correct. Hence, $s_1\sigma \in
L(R)$.

Secondly, the induction method is also used to verify $s \in
L(cl_{i\in I}(S_{i}, \psi_{i})/_{\psi_{fc}}G)$ for any $s \in
L(R)$. (1) $|s|=0$, that is, $s=\epsilon$. We have $\epsilon \in
L(cl_{i\in I}(S_{i}, \psi_{i})/_{\psi_{fc}}G)$. (2) Suppose that
$s \in L(cl_{i\in I}(S_{i}, \psi_{i})/_{\psi_{fc}}G)$ for any $s
\in L(R)$ when $|s|=n$. (3) $|s|=n+1$ with $s=s_1\sigma$. Since
$s_1\sigma \in L(R)$, we have $s_1\sigma \in L(G)$. Then, there is
$x \in \alpha(x_0, s_1)$ such that $x' \in \alpha(x, \sigma)$.
Moreover $|s_1| =n$, we obtain $s_1 \in L(cl_{i\in I}(S_{i},
\psi_{i})/_{\psi_{fc}}G)$. Because $||_{i \in I}S_{i}$ is
deterministic, there is $(y_1, y_2, \cdots, y_n) \in
\beta_{||}(y_{0||}, s_1)$ such that $(x, (y_1, y_2, \cdots, y_n))
\in \alpha'(x_{0}', s_1)$. For $i \in I$, we have $y_i=[s_1]_{i}$
because $s_1\sigma \in L(R)$ and the definition of $\beta_{i}$.
Then, we obtain the following cases. (1) $\sigma \in \Sigma_{uc}$.
Because of $\Sigma_{uci}-compatiblility$ of $\mathcal{S}_{i}$, we
have $\beta_{i}(y_i, \sigma) \neq \emptyset$. Further, $\sigma \in
\psi_{i}(y_i)$ since $\sigma \in \Sigma_{uc}$. Thus, there is
$(y_1', y_2', \cdots, y_n') \in \beta_{||}((y_1, y_2, \cdots,
y_n), \sigma)$ such that $(x', (y_1', y_2', \cdots, y_n')) \in
\alpha'((x, (y_1, y_2, \cdots, y_n)), \sigma)$. Hence, $s_1\sigma
\in L(cl_{i\in I}(S_{i}, \psi_{i})/_{\psi_{fc}}G)$. (2) $\sigma
\in \Sigma_{c}$. If $\sigma \in \Sigma_{uoi}$, then $[s_1]_{i} \in
\beta_{i}([s_1]_{i}, \sigma)$. If $\sigma \notin \Sigma_{uoi}$,
then $[s_1\sigma]_{i} \in \beta_{i}([s_1]_{i}, \sigma)$ because
$s_1\sigma \in L(R)$. Thus, there exists $(y_1', y_2', \cdots,
y_n')\in \beta_{||}((y_1, y_2, \cdots, y_n), \sigma)$. Since
$s_1\sigma \in L(R)$, either $\sigma \notin \Sigma_{ci}$ or
$P_{i}^{-1}P_{i}(s_1)\sigma \cap L(R) \neq \emptyset$. Therefore,
$\sigma \in \cap_{i \in I}\psi_{i}(y_i)=\psi_{fc}$. Then, $(x',
(y_1', y_2', \cdots, y_n')) \in \alpha'((x, (y_1, y_2, \cdots,
y_n)), \sigma)$ which implies $s_1\sigma \in L(cl_{i\in I}(S_{i},
\psi_{i})/_{\psi_{fc}}G)$.

Thirdly, we would like to verify the existence of a bisimulation
relation between the supervised system and the specification.
Because there is a bisimulation relation such that $G||det(R)
\cong_{\phi} R$, we have $L(G||det(R))=L(R)$. In addition, we know
that $L(cl_{i\in I}(S_{i}, \psi_{i})/_{\psi_{fc}}G)=L(R)$. Thus,
$L(cl_{i\in I}(S_{i}, \psi_{i})/_{\psi_{fc}}G)=L(G||det(R))=L(R)$.

Let $\phi_1=\{((x, y), q)~|~\exists s\in L(R)$ s.t. $y \in
\beta_{||}(y_{0||}, s)$, $x \in \alpha(x_0, s)$, $q \in
\delta(q_0, s)$, $z \in \delta_{z}(\{q_0\}, z)$ and $((x, z), q)
\in \phi$\}. For any $((x, y), q) \in \phi_{1}$, if there is a
$\sigma$-successor $(x', y') \in \alpha'((x, y), \sigma)$, where
$\sigma \in \Sigma$, we obtain $s\sigma \in L(cl_{i\in I}(S_{i},
\psi_{i})/_{\psi_{fc}}G)=L(R)$ and $x' \in \alpha(x, \sigma)$.
Thus, there exists $z'\in \delta_{z}(z, \sigma)$ by the definition
of $det(R)$. Then, $(x', z') \in \alpha_{XZ}((x, z), \sigma)$.
Because $((x, z), q) \in \phi$, there exists $q' \in \delta(q,
\sigma)$ such that $((x', z'), q') \in \phi$. Therefore, $((x',
y'), q') \in \phi_1$. If $(x, y) \in X_{m}'$, then $x \in X_{m}$.
It implies $s\in L_{m}(G)$. Because $R$ is marked language closed
with respect to $G$, we have $s \in L_{m}(R)$. Then, $z \in
Z_{m}$. Hence, $(x, z) \in X_{mXZ}$ which implies $q \in Q_{m}$.
For any $(q, (x, y)) \in \phi_1^{-1}$, if there is a
$\sigma$-successor $q'\in \delta(q, \sigma)$, where $\sigma \in
\Sigma$, we have $(x', z') \in \alpha_{XZ}((x, z), \sigma)$ such
that $((x', z'), q') \in \phi$ because $((x, z), q) \in \phi$.
Then, $x' \in \alpha(x, \sigma)$. Further, $s\sigma \in
L(R)=L(cl_{i\in I}(S_{i}, \psi_{i})/_{\psi_{fc}}G)$, there exists
$(x', y')\in \alpha'((x, y), \sigma)$ because of the determinism
of $||_{i \in I}S_{i}$. Hence, $(q', (x', y')) \in \phi_{1}^{-1}$.
If $q \in Q_{m}$, then $x\in X_{m}$. Therefore, $(x, y) \in
X_{m}'$. As a result, $cl_{i\in I}(S_{i}, \psi_{i})/_{\psi_{fc}}G
\cong_{\phi_1 \cup \phi_1^{-1}} R$.
\end{pf}

\begin{remark}\label{conj remark condition}
Intuitively, condition (1) depicts that the nondeterminism of the
plant allowed by the deterministic controller should be equivalent
to the nondeterminism of the desired specification. In addition,
bisimulation implies not only language equivalence but also marked
language equivalence, i.e. $L_{m}(cl_{i\in I}(S_{i},
\psi_{i})/_{\psi_{fc}}G) =L_{m} (R)$, therefore, condition (4) is
required in bisimilarity control. If $R$ is trim, then the
obtained supervisors are nonblocking. Furthermore, condition (1)
always hold when both the plant and the specification are
deterministic. Hence, the decentralized control for language
equivalence \citep{rudie1992think} is a special case of the
decentralized control for bisimulation equivalence.
\end{remark}



\begin{remark}
Except condition (4), which is needed in Theorem 1 because the
marking relies only on the plant in the proposed framework,
conditions (1), (2) and (3) of Theorem 1 for the decentralized
bisimilarity control can be reduced to those are in
\citep{zhoubisimilarity} for the the centralized bisimilarity
control when $n=1$. Therefore, the result for the centralized
framework of bisimilarity control is a special case for the
decentralized framework of bisimilarity control in this paper.
\end{remark}


\begin{remark}\label{conj remark sup}
From the sufficiency part of Theorem \ref{conjunctive}, it is
shown that decentralized bisimilarity supervisors can be designed
according to (\ref{con sup}) and (\ref{con sup dec}) when the
necessary and sufficient condition has been satisfied.
\end{remark}

\begin{remark}\label{conj remark complexity}
To obtain the computational complexity of verifying the existence
condition of decentralized bisimilarity supervisors for the
conjunctive architecture, we examine the conditions of Theorem
\ref{conjunctive} item by item. (1) $G||det(R)\cong_{\phi} R$.
Since both the plant and the specification are nondeterministic,
their numbers of transitions are $O(|X|^{2}|\Sigma|)$ and
$O(|Q|^{2}|\Sigma|)$ respectively. Moreover, $det(R)$ is
deterministic with $O(2^{|Q|}|\Sigma|)$ transitions. According to
\citep{fernandez1990implementation}, the complexity of checking $G
|| det(R) \cong_{\phi} R$ is
$O(|X|^{2}2^{|Q|^{2}}|\Sigma|log(|X|2^{|Q|}))$. (2) $L(R)$ is
language controllable with respect to $L(G)$ and $\Sigma_{uc}$,
which can be tested with complexity $O(|X|^{2}|Q|^{2}|\Sigma|)$
\citep{cassandras2008introduction}. (3) $L(R)$ is $C\&P$
co-observable with respect to $L(G)$, $\Sigma_{ci}$ and
$\Sigma_{oi}$, where $i\in I$. It can be verified by polynomial
complexity with respect to $|X|$ and $|Q|$. (4) $R$ is marked
language closed with respect to $G$. By checking the states of
$G||R$, the condition (4) can be tested with complexity
$O(|X||Q|)$. Therefore, the computational complexity of verifying
the conditions of Theorem \ref{conjunctive} is
$O(|X|^{2}2^{|Q|^{2}}|\Sigma|log(|X|2^{|Q|}))$, which is
exponential with respect to $|X|$ and $|Q|$.
\end{remark}

\section{Disjunctive Architecture}
In this section, we study the decentralized bisimilarity control
under the disjunctive architecture. As below, the existence result
for the disjunctive architecture is presented.

\begin{theorem}\label{disjunctive}
Given a plant $G=(X, \Sigma, \alpha, x_0, X_{m})$ and a
specification $R = (Q, \Sigma, \delta, q_0, Q_{m})$ with $L(R)
\subseteq L(G)$, there exist decentralized $(\Sigma_{uoi},
\Sigma_{uci})-compatible$ supervisors $\mathcal{S}_{i}=(S_{i},
\psi_{i})$ with the disjunctive decision fusion rule $\psi_{fd}$
such that $cl_{i\in I}(S_{i}, \psi_{i})/_{\psi_{fd}}G \cong R$ if
the following conditions hold:

\begin{enumerate}

\item[(1)] There is a bisimulation relation $\phi$ such that $G ||
det(R) \cong_{\phi} R$;

\item[(2)] $L(R)$ is language controllable with respect to $L(G)$
and $\Sigma_{uc}$;

\item[(3)] $L(R)$ is $D\&A$ co-observable with respect to $L(G)$,
$\Sigma_{ci}$ and $\Sigma_{oi}$, where $i \in I$.

\item[(4)] $R$ is marked language closed with respect to $G$.
\end{enumerate}

\end{theorem}

\begin{pf}
Let $det(R)=(Z, \Sigma, \delta_{Z}, \{q_0\}, Z_{m})$ and
$G||det(R)=(X_{XZ}, \Sigma, \alpha_{XZ}, (x_0, \{q_0\}),
X_{mXZ})$.

(Necessity) Consider $\mathcal{S}_{i}=(S_{i}, \psi_{i})=((Y_{i},
\Sigma, \beta_{i}, y_{0i}, Y_{mi}), \psi_{i})$ and $||_{i \in
I}S_i=(Y_{||}, \Sigma, \beta_{||}, y_{0||}, Y_{m||})$, where $i
\in I$. Because there is a bisimulation relation $\phi'$ such that
$cl_{i\in I}(S_{i}, \psi_{i})/_{\psi_{fd}}G \cong_{\phi'} R$. We
have $L(cl_{i\in I}(S_{i}, \psi_{i})/_{\psi_{fd}}G) =L(R)$ and
$L_{m}(cl_{i\in I}(S_{i}, \psi_{i})/_{\psi_{fd}}G)=L_{m}(R)$.


Firstly, we prove that $L(R)$ is language controllable with
respect to $L(G)$ and $\Sigma_{uc}$. For any $s \in L(R)$ and
$\sigma \in \Sigma_{uc}$ such that $s\sigma \in L(G)$, there is $x
\in \alpha(x_0, s)$ with $x' \in \alpha(x, \sigma)$. Because $s
\in L(R)=L(cl_{i\in I}(S_{i}, \psi_{i})/_{\psi_{fd}}G)$ and $||_{i
\in I}S_{i}$ is deterministic, there exists $(x, (y_1, y_2,
\cdots, y_n)) \in \alpha'(x_{0}', s)$. Because $\sigma \in
\Sigma_{uc}$, we have $\sigma \in \psi_{fd}(y_1, y_2, \cdots,
y_n)$. Moreover, $\mathcal{S}_{i}$ is $\Sigma_{uci}-compatible$,
which implies $\beta_{i}(y_i, \sigma) \neq \emptyset$ for $i \in
I$. Thus, there is $(y_1', y_2', \cdots, y_n') \in
\beta_{||}((y_1, y_2, \cdots, y_n), \sigma)$ such that $(x',
(y_1', y_2', \cdots, y_n')) \in \alpha'((x, (y_1, y_2, \cdots,
y_n)), \sigma)$ according to Definition \ref{supvised system}.
Therefore, $s\sigma \in L(cl_{i\in I}(S_{i},
\psi_{i})/_{\psi_{fd}}G)=L(R)$.

Secondly, we verify $D\&A$ co-observability of $L(R)$ with respect
to $L(G)$, $\Sigma_{ci}$ and $\Sigma_{oi}$ for $i \in I$. Assume
that there is $s \in L(R)$ and $\sigma \in \Sigma_{c}$ satisfying
$s\sigma \in L(R)$, moreover, either $\sigma \notin \Sigma_{ci}$
or $(P_i^{-1}P_i(s) \cap L(R))\sigma \cap L(G) \nsubseteq L(R)$
for any $i \in I$. Then, either $\sigma \notin \Sigma_{ci}$ or
there is $s' \in \Sigma^{*}$ such that $P_{i}(s)=P_{i}(s')$ and
$s'\sigma \in L(G)\setminus L(R)$ for $i \in I$. Because $s\sigma
\in L(R)$, we have $s\sigma \in L(cl_{i\in I}(S_{i},
\psi_{i})/_{\psi_{fd}}G)$. Then, there exists $(x, (y_1, y_2
\cdots, y_n)) \in \alpha'(x_{0}', s)$ such that $(x', (y_1', y_2'
\cdots, y_n')) \in \alpha'((x, (y_1, y_2 \cdots, y_n)), \sigma)$.
According to Definition \ref{supvised system} and (\ref{disj
global dec}), $(y_1', y_2' \cdots, y_n') \in \beta_{||}((y_1, y_2
\cdots, y_n), \sigma)$ and there exists $i \in I$ such that
$\sigma \in \Sigma_{ci}$ and $\sigma \in \psi_{i}(y_i)$. Because
$s'\sigma \in L(G)$ and $s' \in L(R)=L(cl_{i\in I}(S_{i},
\psi_{i})/_{\psi_{fd}}G$, there are $x'' \in \alpha(x_0, s')$ and
$(y_1'', y_2'', \cdots, y_n'') \in \beta_{||}(y_{0||}, s')$ such
that $(x'', (y_1'', y_2'', \cdots, y_n'')) \in \alpha'(x_{0}',
s')$ and $x''' \in \alpha(x'', \sigma)$. For $i \in I$, if $\sigma
\in \Sigma_{ci}$, we have $P_{i}(s)=P_{i}(s')$ with $y_i=y_i''$
because of $\Sigma_{uoi}-compability$ of $\mathcal{S}_{i}$.
Therefore, either $\sigma \notin \Sigma_{ci}$ or $y_i=y_i''$ for
$i \in I$. It implies $\sigma \in \cup_{i \in I}\psi_{i}(y_i'')$.
Furthermore, $\mathcal{S}_{i}$ is $\Sigma_{uci}-compatible$. Then,
there is $(y_1''', y_2''', \cdots, y_n''') \in \beta_{||}((y_1'',
y_2'', \cdots, y_n''), \sigma)$ such that $(x''', (y_1''', y_2''',
\cdots, y_n''')) \in \alpha'((x'', (y_1'', y_2'', \cdots, y_n'')),
\sigma)$. It implies $s'\sigma \in L(cl_{i\in I}(S_{i},
\psi_{i})/_{\psi_{fd}}G)=L(R)$, which contradicts that $s'\sigma
\notin L(R)$. Therefore, the assumption is not correct. It implies
$L(R)$ is $D\&A$ co-observable with respect to $L(G)$ and
$\Sigma_{ci}$ and $\Sigma_{oi}$.

Thirdly, we would like to prove that there is a bisimulation
relation $\phi$ such that $G || det(R) \cong_{\phi} R$. From the
definition of product, we have $L(G||det(R))=L(G) \cap
L(det(R))=L(R)$. Thus, $L(cl_{i\in I}(S_{i},
\psi_{i})/_{\psi_{fd}}G)=L(R)=L(G||det(R))$. Let $\phi_1=\{((x,
z), q) \in X_{XZ} \times Q~|~\exists s\in L(R)$ s.t. $(x, z) \in
\alpha_{XZ}((x_0, \{q_0\}), s)$, $q \in \delta(q_0, s)$, $y \in
\beta_{||}(y_{0||}, s)$ and $((x, y), q) \in \phi'$\}. For any
$((x, z), q) \in \phi_1$, if there is a $\sigma$-successor $(x',
z') \in \alpha_{XZ}((x, z), \sigma)$, where $\sigma \in \Sigma$,
we obtain $s\sigma \in L(R)=L(cl_{i\in I}(S_{i},
\psi_{i})/_{\psi_{fd}}G)$ and $x' \in \alpha(x, \sigma)$. Since
$||_{i\in I}S_{i}$ is deterministic, there is $y \in
\beta_{||}(y_{0||}, s)$ such that $y' \in \beta_{||} (y, \sigma)$.
It implies $(x', y') \in \alpha'((x, y), \sigma)$. Then, there
exists $q' \in \delta(q, \sigma)$ such that $((x', y'), q') \in
\phi'$. Hence, $((x', z'), q') \in \phi_{1}$. If $(x, z) \in
X_{mXZ}$, then $x \in X_{m}$, which implies $(x, y) \in X_{m}'$.
Therefore, $q \in Q_{m}$. For any $(q, (x, z)) \in \phi_1^{-1}$,
if there is a $\sigma$-successor $q'\in \delta(q, \sigma)$, where
$\sigma \in \Sigma$, we have $(x', y') \in \alpha'((x, y),
\sigma)$ such that $((x', y'), q') \in \phi'$ because $((x, y), q)
\in \phi'$. Further, $s\sigma \in L(R)$, there exists $z' \in
\delta_{Z}(z, \sigma)$ by the definition of $det(R)$. Thus, $(x',
z') \in \alpha_{XZ}((x, z), \sigma)$. Hence, $(q', (x', z')) \in
\phi_{1}^{-1}$. If $q \in Q_{m}$, then $z \in Z_{m}$ and $x\in
X_{m}$. Therefore, $(x, z) \in X_{mXZ}$. As a result, $G||det(R)
\cong_{\phi_1 \cup \phi_1^{-1}} R$.

Similar to Theorem \ref{conjunctive}, we can also prove that $R$
is marked language closed with respect to $G$.




(Sufficiency) We construct $\mathcal{S}_{i}=(S_{i},
\psi_{i})=((Y_{i}, \Sigma, \beta_{i}, y_{0i}, Y_{mi}), \psi_{i})$
as follows. The automaton $S_{i}$ is as the same as (\ref{con
sup}) and for any $y_{i} \in Y_{i}$, the local decision map
$\psi_{i}(y_i)$ is defined as:

\begin{equation}\label{dis sup dec}
\psi_{i}(y_i) = \left\{ {\begin{array}{*{20}c}
 \Sigma_{uc}
\cup \{\sigma \in \Sigma_{ci}~|~([s]_{i}\cap
L(R)) \sigma \\ \cap L(G) \subseteq L(R)\} & {y_{i}=[s]_{i} };  \\ 
\Sigma_{uc} & {y_i = z_{di} }.  \\
\end{array}} \right.
\end{equation}

It can be seen that $\mathcal{S}_{i}$ is
$(\Sigma_{uoi},\Sigma_{uci})-compatible$ and $\psi_{i}$ meets the
requirement of the disjunctive architecture. Let $\psi_{fd}$
(\ref{disj global dec}) be the disjunctive decision fusion rule.

Firstly, we prove that $s \in L(R)$ for any $s \in L(cl_{i\in
I}(S_{i}, \psi_{i})/_{\psi_{fd}}G)$ by the induction method. (1)
$|s|=0$, that is, $s=\epsilon$. We have $\epsilon \in L(R)$. (2)
Suppose that $s \in L(R)$ for any $s \in L(cl_{i\in I}(S_{i},
\psi_{i})/_{\psi_{fd}}G)$ when $|s|=n$. (3) $|s|=n+1$ with
$s=s_1\sigma$. Assume that $s_1\sigma \notin L(R)$. Since
$s_1\sigma \in L(cl_{i\in I}(S_{i}, \psi_{i})/_{\psi_{fd}}G)$,
there is $(x, (y_1, y_2, \cdots, y_n)) \in \alpha'(x_0', s_1)$ and
$\sigma \in \Sigma$ such that $(x', (y_1', y_2', \cdots, y_n'))
\in \alpha'((x, (y_1, y_2, \cdots, y_n)), \sigma)$. Then,
$s_1\sigma \in L(G)$ and $\sigma \in \cup_{i\in
I}\psi_{i}(y_i)=\psi_{fd}(y_1, y_2, \cdots, y_n)$. We have the
following cases. Case 1: $\sigma \in \Sigma_{uc}$. Because
$|s_1|=n$, we have $s_1 \in L(R)$. Then, $s_1\sigma \in L(R)$
since  $s_1\sigma \in L(G)$ and $L(R)$ is language controllable
with respect to $L(G)$ and $\Sigma_{uc}$. Thus, there is a
contradiction. Case 2: $\sigma \in \Sigma_{c}$.  Since $s_1 \in
L(R)$, we obtain $y_i=[s_1]_{i}$ for $i \in I$ by the definition
of $\beta_{i}$ and $s_1 \in L(R)$. Because $\sigma \in \cup_{i\in
I}\psi_{i}(y_i)$, there is $i \in I$ such that $([s_1]_{i}\cap
L(R))\sigma \cap L(G) \subseteq L(R)$. Therefore, $s_1\sigma \in
L(R)$, which introduces a contradiction. Then, the assumption is
not correct. Hence, $s_1\sigma \in L(R)$.

Secondly, the induction method is also used to prove that $s \in
L(cl_{i\in I}(S_{i}, \psi_{i})/_{\psi_{fd}}G)$ for any $s \in
L(R)$. (1) $|s|=0$, that is, $s=\epsilon$. We have $\epsilon \in
L(cl_{i\in I}(S_{i}, \psi_{i})/_{\psi_{fd}}G)$. (2) Suppose that
$s \in L(cl_{i\in I}(S_{i}, \psi_{i})/_{\psi_{fd}}G)$ for any $s
\in L(R)$ when $|s|=n$. (3) $|s|=n+1$ with $s=s_1\sigma$. Since
$s_1\sigma \in L(R)$, we have $s_1\sigma \in L(G)$. Then, there is
$x \in \alpha(x_0, s_1)$ such that $x' \in \alpha(x, \sigma)$.
Moreover $|s_1|=n$, we obtain that $s_1 \in L(cl_{i\in I}(S_{i},
\psi_{i})/_{\psi_{fd}}G$. Because $||_{i \in I}S_{i}$ is
deterministic, there is $(y_1, y_2, \cdots, y_n) \in
\beta_{||}(y_{0||}, s_1)$ such that $(x, (y_1, y_2, \cdots, y_n))
\in \alpha'(x_{0}', s_1)$. For $i \in I$, we have $y_i=[s_1]_{i}$
because $s_1\sigma \in L(R)$ and the definition of $\beta_{i}$.
Then, we obtain the following cases. (1) $\sigma \in \Sigma_{uc}$.
Because $\mathcal{S}_{i}$ is $\Sigma_{uci}-compatible$, we obtain
that $\beta_{i}(y_i, \sigma) \neq \emptyset$ and $\sigma \in
\psi_{i}(y_i)$. It implies there is $(y_1', y_2', \cdots, y_n')\in
\beta_{||}((y_1, y_2, \cdots, y_n), \sigma)$ such that $(x',
(y_1', y_2', \cdots, y_n')) \in \alpha'((y_1, y_2, \cdots, y_n),
\sigma)$. Hence, $s_1\sigma \in L(cl_{i\in I}(S_{i},
\psi_{i})/_{\psi_{fd}}G)$. (2) $\sigma \in \Sigma_{c}$. If $\sigma
\in \Sigma_{uoi}$, then $[s_1]_{i} \in \beta_{i}([s_1]_{i},
\sigma)$. If $\sigma \notin \Sigma_{uoi}$, then $[s_1\sigma]_{i}
\in \beta_{i}([s_1]_{i}, \sigma)$ because $s_1\sigma \in L(R)$.
Thus, there is $(y_1', y_2', \cdots, y_n')\in \beta_{||}((y_1,
y_2, \cdots, y_n), \sigma)$. Since $s_1\sigma \in L(R)$ and $D\&A$
co-observability of $L(R)$ with respect to $L(G)$, $\Sigma_{oi}$
and $\Sigma_{ci}$, there exists $i \in I$ such that $\sigma \in
\Sigma_{ci}$ and $(P_{i}^{-1}P_{i}(s_1)\cap L(R))\sigma \cap L(G)
\subseteq L(R)$. It implies $\sigma \in \psi_{i}(y_i)$. Therefore,
$\sigma \in \cup_{i \in I}\psi_{i}(y_i)=\psi_{fd}$. Then, $(x',
(y_1', y_2', \cdots, y_n')) \in \alpha'((x, (y_1, y_2, \cdots,
y_n)), \sigma)$, which implies $s_1\sigma \in L(cl_{i\in I}(S_{i},
\psi_{i})/_{\psi_{fd}}G)$.

Thirdly, we prove that there exists a bisimulation relation
between the supervised system and the specification. Because there
is a bisimulation relation such that $G||det(R) \cong_{\phi} R$,
we have $L(G||det(R))=L(R)$. In addition, we know $L(cl_{i\in
I}(S_{i}, \psi_{i})/_{\psi_{fd}}G)=L(R)$ Thus, $L(cl_{i\in
I}(S_{i}, \psi_{i})/_{\psi_{fd}}G)=L(G||det(R))=L(R)$. Let
$\phi_1=\{((x, y), q)~|~\exists s\in L(R)$ s.t. $y \in
\beta_{||}(y_{0||}, s)$, $x \in \alpha(x_0, s)$, $q \in
\delta(q_0, s)$, $z \in \delta_{z}(\{q_0\}, z)$ and $((x, z), q)
\in \phi$\}. Similar to Theorem \ref{conjunctive}, we can obtain
that $cl_{i\in I}(S_{i}, \psi_{i})/_{\psi_{fd}}G \cong_{\phi_1
\cup \phi_1^{-1}} R$.


%

\end{pf}


\begin{remark}\label{disj remark sup}
When the conditions of Theorem \ref{disjunctive} hold, we can
construct decentralized bisimilarity supervisors for the
disjunctive architecture by (\ref{con sup}) and (\ref{dis sup
dec}), which is proved in the sufficiency part of Theorem
\ref{disjunctive}.
\end{remark}

\begin{remark}\label{disj remark complexity}
From \citep{yoo2002general}, $D\&A$ co-observability of $L(R)$
with respect to $L(G)$, $\Sigma_{ci}$ and $\Sigma_{oi}$, where
$i\in I$, can be verified by polynomial complexity with respect to
$|X|$ and $|Q|$. According to Remark \ref{conj remark complexity},
the computational complexity of verifying the existence condition
of decentralized bisimilarity supervisors for the disjunctive
architecture is $O(|X|^{2}2^{|Q|^{2}}|\Sigma|log(|X|2^{|Q|}))$,
which is exponential with respect to $|X|$ and $|Q|$.
\end{remark}

\section{General Architecture}
Under the general architecture, we explore the decentralized
bisimilarity control in this section. Then, the following Theorem
\ref{general} depicts the necessary and sufficient condition for
the existence of decentralized bisimilarity supervisors for the
general architecture.

\begin{theorem}\label{general}
Given a plant $G=(X, \Sigma, \alpha, x_0, X_{m})$ and a
specification $R = (Q, \Sigma, \delta, q_0, Q_{m})$ with $L(R)
\subseteq L(G)$, there is $(\Sigma_{uoi},
\Sigma_{uci})-compatible$ supervisors $\mathcal{S}_{i}=(S_{i},
\psi_{i})$ with the general decision fusion rule $\psi_{fg}$ such
that $cl_{i\in I}(S_{i}, \psi_{i})/_{\psi_{fg}}G \cong R$ if and
only if the following conditions hold:

\begin{enumerate}
\item[(1)] There is a bisimulation relation $\phi$ such that $G ||
det(R) \cong_{\phi} R$;

\item[(2)] $L(R)$ is language controllable with respect to $L(G)$
and $\Sigma_{uc}$;

\item[(3)] $L(R)$ is said to be co-observable with respect to
$L(G)$, $\Sigma_{oi}$, $\Sigma_{cei}$ and $\Sigma_{cdi}$, where
$i\in I$;


\item[(4)] $R$ is marked language closed with respect to $G$.
\end{enumerate}
\end{theorem}

\begin{pf}
Consider $det(R)=(Z, \Sigma, \delta_{Z}, \{q_0\}, Z_{m})$ and
$G||det(R)=(X_{XZ}, \Sigma, \alpha_{XZ}, (x_0, \{q_0\}),
X_{mXZ})$.

(Necessity) Similar to Theorem \ref{conjunctive} and Theorem
\ref{disjunctive}, we can prove that the necessity part of this
theorem.

(Sufficiency) The local supervisors $\mathcal{S}_{i}=(S_{i},
\psi_{i})=((Y_{i}, \Sigma, \beta_{i}, y_{0i}, Y_{mi}), \psi_{i})$
is designed as follows. The automaton $S_{i}$ is as the same as
(\ref{con sup}) and for any $y_{i} \in Y_{i}$, the local decision
map $\psi_{i}(y_i)$ is defined as:

\begin{equation}\label{gen sup dec}
\psi_{i}(y_i) = \left\{ {\begin{array}{*{20}c}
 \Sigma_{uc} \cup \Sigma_{ce}\setminus
\Sigma_{cei} \cup \{\sigma \in \Sigma_{cei}~|~\exists s' \in
[s]_{i},  s'\sigma \in L(R)\} \cup  \\
\{\sigma \in \Sigma_{cdi}~|~([s]_{i} \cap L(R))\sigma \cap
L(G) \subseteq L(R)\} & {y_{i}=[s]_{i} };  \\ 
\Sigma_{uc} \cup \Sigma_{ce}\setminus \Sigma_{cei} & {y_i = z_{di} }.  \\
\end{array}} \right.
\end{equation}

It can be seen that $\mathcal{S}_{i}$ is
$(\Sigma_{uoi},\Sigma_{uci})-compatible$ and $\psi_{i}$ meeting
the requirement of the general architecture. Let $\psi_{fg}$
(\ref{general global dec}) be the generalized decision fusion
rule.

Next, we would like to prove $L(R)=L(cl_{i\in I}(S_{i},
\psi_{i})/_{\psi_{fg}}G)$. Firstly, we verify that $s \in L(R)$
for any $s \in L(cl_{i\in I}(S_{i}, \psi_{i})/G)$ by the induction
method. (1) $|s|=0$, that is, $s=\epsilon$. We have $\epsilon \in
L(R)$. (2) Suppose that $s \in L(R)$ for any $s \in L(cl_{i\in
I}(S_{i}, \psi_{i})/_{\psi_{fg}}G)$ when $|s|=n$. (3) $|s|=n+1$
with $s=s_1\sigma$. Assume that $s_1\sigma \notin L(R)$. Since
$s_1\sigma \in L(cl_{i\in I}(S_{i}, \psi_{i})/_{\psi_{fg}}G)$,
there is $(x, (y_1, y_2, \cdots, y_n)) \in \alpha'(x_0', s_1)$ and
$\sigma \in \Sigma$ such that $(x', (y_1', y_2', \cdots, y_n'))
\in \alpha'((x, (y_1, y_2, \cdots, y_n)), \sigma)$. Then
$s_1\sigma \in L(G)$ and $\sigma \in \psi_{fg}(y_1, y_2, \cdots,
y_n)$. We have the following cases. Case 1: $\sigma \in
\Sigma_{uc}$. Then, $s_1\sigma \in L(R)$ because of the language
controllability of $L(R)$ with respect to $L(G)$ and
$\Sigma_{uc}$. Thus, there is a contradiction. Case 2: $\sigma \in
\Sigma_{ce}$. Then, $\sigma \in P_{ce}(\cap_{i\in
I}\psi_{i}(y_i))$. Since $s_1 \in L(R)$, we obtain $y_i=[s_1]_{i}$
for $i \in I$ by the definition of $\beta_{i}$ and $s_1 \in L(R)$.
Thus, either $\sigma \notin \Sigma_{cei}$ or there is $s_1' \in
[s_1]_{i}$ such that $s_1'\sigma \in L(R)$. That is,
$P_{i}^{-1}P_{i}(s_1)\sigma \cap L(R) \neq \emptyset$. Therefore,
$L(R)$ is not $C\&P$ co-observability of with respect to $L(G)$,
$\Sigma_{oi}$ and $\Sigma_{cei}$, where $i \in I$. It introduces a
contradiction. Hence, the assumption is not correct. As a result,
$s_1\sigma \in L(R)$. Case 3: $\sigma \in \Sigma_{cd}$. Thus,
$\sigma \in P_{cd}(\cup_{i\in I}\psi_{i}(y_i))$. Because $s_1 \in
L(R)$, we obtain $y_i=[s_1]_{i}$ for $i \in I$ by the definition
of $\beta_{i}$ and $s_1 \in L(R)$. Because $\sigma \in
P_{cd}(\cup_{i\in I}\psi_{i}(y_i))$, there is $i \in I$ such that
$([s_1]_{i})\sigma \cap L(G) \subseteq L(R)$. It implies
$s_1\sigma \in L(R)$. Then, there exists a contradiction.
Therefore, the assumption is not correct, which implies $s_1\sigma
\in L(R)$.

As below, the induction method is also used to prove $s \in
L(cl_{i\in I}(S_{i}, \psi_{i})/_{\psi_{fg}}G)$ for any $s \in
L(R)$. (1) $|s|=0$, that is, $s=\epsilon$. We have $\epsilon \in
L(cl_{i\in I}(S_{i}, \psi_{i})/_{\psi_{fg}}G)$. (2) Suppose that
$s \in L(cl_{i\in I}(S_{i}, \psi_{i})/_{\psi_{fg}}G)$ for any $s
\in L(R)$ when $|s|=n$. (3) $|s|=n+1$ with $s=s_1\sigma$. Since
$s_1\sigma \in L(R)$, we have $s_1\sigma \in L(G)$. Then, there is
$x \in \alpha(x_0, s_1)$ such that $x' \in \alpha(x, \sigma)$.
Moreover $|s_1| =n$, we obtain $s_1 \in L(cl_{i\in I}(S_{i},
\psi_{i})/_{\psi_{fg}}G)$. Because $||_{i \in I}S_{i}$ is
deterministic, there is $(y_1, y_2, \cdots, y_n) \in
\beta_{||}(y_{0||}, s_1)$ such that $(x, (y_1, y_2, \cdots, y_n))
\in \alpha'(x_{0}', s_1)$. For $i \in I$, we have $y_i=[s_1]_{i}$
because $s_1\sigma \in L(R)$ and the definition of $\beta_{i}$.
Then, we obtain the following cases. Case 1: $\sigma \in
\Sigma_{uc}$. Because of $\Sigma_{uci}-compatiblility$ of
$\mathcal{S}_{i}$, we have $\beta_{i}(y_i, \sigma) \neq
\emptyset$. Further, $\sigma \in \psi_{i}(y_i)$ since $\sigma \in
\Sigma_{uc}$. It implies there is $(y_1', y_2', \cdots, y_n') \in
\beta_{||}((y_1, y_2, \cdots, y_n), \sigma)$ such that $(x',
(y_1', y_2', \cdots, y_n')) \in \alpha'((x, (y_1, y_2, \cdots,
y_n)), \sigma)$. Hence, $s_1\sigma \in L(cl_{i\in I}(S_{i},
\psi_{i})/_{\psi_{fg}}G)$.  Case 2: $\sigma \in \Sigma_{ce}$. If
$\sigma \in \Sigma_{uoi}$, then $[s_1]_{i} \in
\beta_{i}([s_1]_{i}, \sigma)$. If $\sigma \notin \Sigma_{uoi}$,
then $[s_1\sigma]_{i} \in \beta_{i}([s_1]_{i}, \sigma)$ because
$s_1\sigma \in L(R)$. Thus, there is $(y_1', y_2', \cdots,
y_n')\in \beta_{||}((y_1, y_2, \cdots, y_n), \sigma)$. Since
$s_1\sigma \in L(R)$, either $\sigma \in \Sigma_{cei}$ or
$P_{i}^{-1}P_{i}(s_1)\sigma \cap L(R) \neq \emptyset$. Therefore,
$\sigma \in P_{ce}(\cap_{i \in I}\psi_{i}(y_i))$. Then, $(x',
(y_1', y_2', \cdots, y_n')) \in \alpha'((x, (y_1, y_2, \cdots,
y_n)), \sigma)$, which implies $s_1\sigma \in L(cl_{i\in I}(S_{i},
\psi_{i})/_{\psi_{fg}}G)$.  Case 3: $\sigma \in \Sigma_{cd}$.
Similar to Case 2, we can prove that there is $(y_1', y_2',
\cdots, y_n')\in \beta_{||}((y_1, y_2, \cdots, y_n), \sigma)$.
Since $s_1\sigma \in L(R)$ and $D\&A$ co-observability of $L(R)$
with respect to $L(G)$, $\Sigma_{oi}$ and $\Sigma_{cdi}$, there
exists $i \in I$ such that $\sigma \in \Sigma_{cdi}$ and
$(P_{i}^{-1}P_{i}(s_1)\cap L(R))\sigma \cap L(G) \subseteq L(R)$.
It implies $\sigma \in P_{cd}(\cup_{i\in I}\psi_{i}(y_i))$. Then,
$(x', (y_1', y_2', \cdots, y_n')) \in \alpha'((x, (y_1, y_2,
\cdots, y_n)), \sigma)$, which implies $s_1\sigma \in L(cl_{i\in
I}(S_{i}, \psi_{i})/_{\psi_{fg}}G)$.

Because there is a bisimulation relation such that $G||det(R)
\cong_{\phi} R$, we have $L(G||det(R))=L(R)$. In addition, we know
$L(cl_{i\in I}(S_{i}, \psi_{i})/_{\psi_{fg}}G)=L(R)$. Thus,
$L(cl_{i\in I}(S_{i}, \psi_{i})/_{\psi_{fg}}G)=L(G||det(R))=L(R)$.
Let $\phi_1=\{((x, y), q)~|~\exists s\in L(R)$ s.t. $y \in
\beta_{||}(y_{0||}, s)$, $x \in \alpha(x_0, s)$, $q \in
\delta(q_0, s)$, $z \in \delta_{z}(\{q_0\}, z)$ and $((x, z), q)
\in \phi$\}. It can be easily obtained $cl_{i\in I}(S_{i},
\psi_{i})/_{\psi_{fg}}G \cong_{\phi_1 \cup \phi_1^{-1}} R$.


%
\end{pf}

\begin{figure}[!htb]
\begin{center}
\includegraphics*[scale=.30]{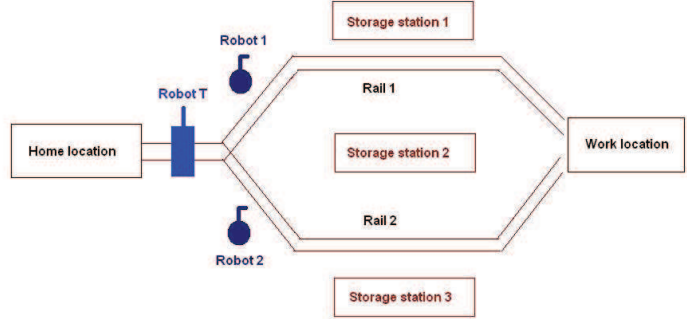}
\caption{Manufacturing System} \label{manu}
\end{center}
\end{figure}

\begin{remark}\label{gen remark condition}
Since $C\&P$ co-observability and $D\&A$ co-observability are the
special cases of co-observability, the result of Theorem
\ref{general} for the general architecture generalizes the results
of Theorem \ref{conjunctive} and Theorem \ref{disjunctive}.
Moreover, decentralized bisimilarity supervisors can be designed
by using (\ref{con sup}) and (\ref{gen sup dec}) for the general
architecture.
\end{remark}

%

\begin{figure}[!htb]
\begin{center}
\includegraphics*[scale=.70]{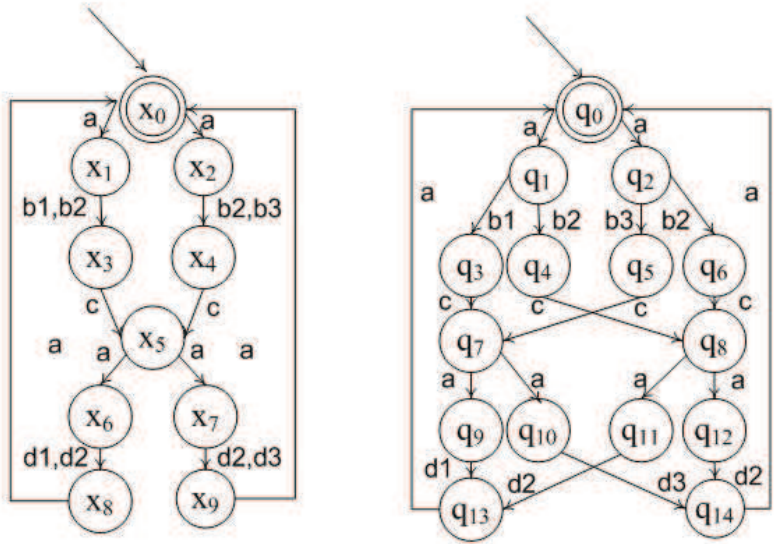}
\caption{Plant $G$ (Left) and Specification $R$ (Right) of Example
1} \label{cp}
\end{center}
\end{figure}

\begin{remark}\label{gen remark complexity}
Refer to Remark \ref{conj remark complexity} and Remark \ref{disj
remark complexity}, the computational complexity of verifying the
existence condition of decentralized bisimilarity supervisors for
the general architecture is
$O(|X|^{2}2^{|Q|^{2}}|\Sigma|log(|X|2^{|Q|}))$, which is
exponential with respect to $|X|$ and $|Q|$.
\end{remark}

\section{Illustrative Examples}
In this section, four examples are provided to demonstrate the
proposed results.



\begin{example}
Consider the following manufacturing example adopted from
\citep{zhou2007small}. A manufacturing system consists of a home
location, a work location, three storage stations and three
robots, which is shown in Fig. \ref{manu}. Robot $T$ is available
at its home location to traverse on one of the two rails.
Traversal on Rail $i$ $(i = 1, 2)$ is randomly chosen and is
denoted by event $a$. While Robot $T$ is on Rail $i$, it can pick
a part from Storage $i$ (event $b_{i}$) or Storage $(i+1)$ (event
$b_{i+1}$), and then it takes the part to work location for
processing (event $c$). When returning, Robot $T$ can
nondeterministically choose a Rail-$i$ and drop the part to either
Storage $i$ (event $d_i$) or Storage $(i + 1)$ (event $d_{i+1}$)
and returns to its home location. Robot $1$ and Robot $2$ can
monitor and supervise the manufacturing process.


The control specification requires that a part be returned to its
original pickup location except the parts picked up at Storage 1
(respectively Storage 3) can also be returned to Storage 3
(respectively, Storage 1), as those parts are exchangeable. The
specification also requires that Robot $T$ always be able to
return to its home location (which means that the state
representing the home location is the only marked state). Models
$G$ and $R$ of the manufacturing system and its specification are
given in Fig. \ref{cp}.
\end{example}

\begin{figure}[!htb]
\begin{center}
\includegraphics*[scale=.70]{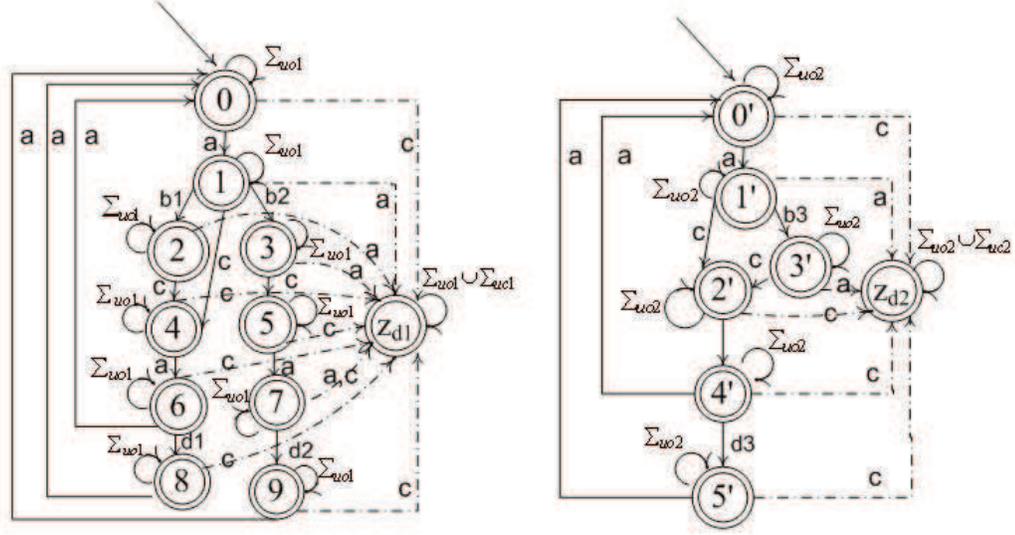}
\caption{The Automata of $\mathcal{S}_{1}$ (Left) and
$\mathcal{S}_{2}$ (Right) of Example 1} \label{clsup}
\end{center}
\end{figure}

\begin{figure}[!htb]
\begin{center}
\includegraphics*[scale=.70]{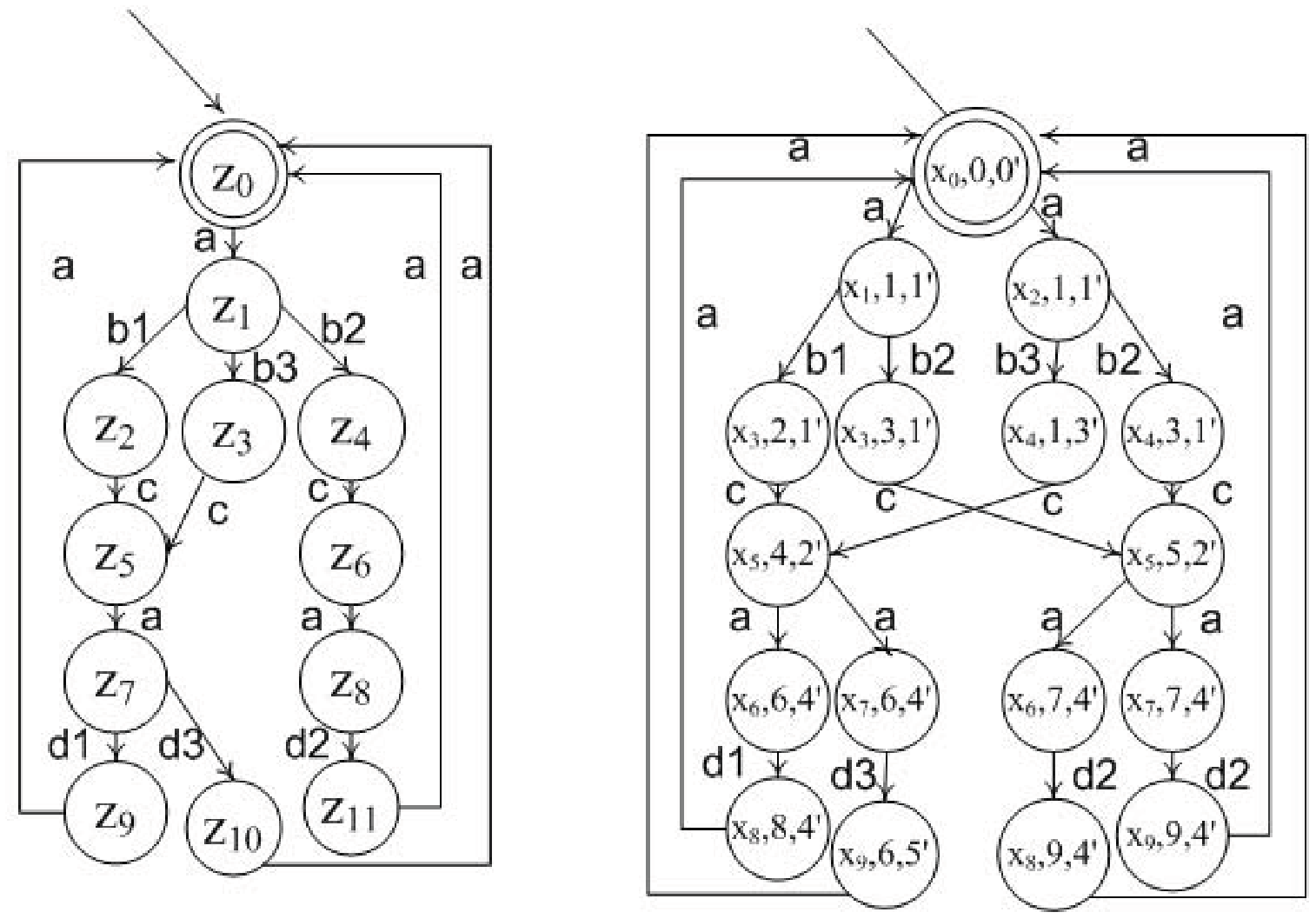}
\caption{det(R) (Left) and Supervised System $cl(\mathcal{S}_1,
\mathcal{S}_2)/_{\psi_{fc}}G$ (Right) of Example 1} \label{ccls}
\end{center}
\end{figure}

Suppose $\Sigma_{o1}=\{a, c, b_1, b_2, d_1, d_2\}$,
$\Sigma_{o2}=\{a, c, b_3, d_3\}$, $\Sigma_{c1}=\{b_1, b_2, d_1,
d_2, d_3\}$ and $\Sigma_{c2}=\{b_3, d_3\}$. Then,
$\Sigma_{uc}=\{a, c \}$, $\Sigma_{uc1}=\{a, c, b_3\}$ and
$\Sigma_{uc2}=\{a, c, b_1, d_1, b_2, d_2\}$. For this example, we
obtain that
$L(G)=\overline{(ab_{1}cad_{1}a+ab_{1}cad_{2}a+ab_{1}cad_{3}a+ab_{2}}$\\
$\overline{cad_{1}a+ab_{2}cad_{2}a+ab_{2}cad_{3}a+ab_{3}cad_{1}a+ab_{3}cad_{2}a+ab_{3}cad_{3}a)^{*}}$
and
$L(R)=\overline{(a}$\\$\overline{b_{1}cad_{1}a+ab_{1}cad_{3}a+ab_{2}cad_{2}a+ab_{3}cad_{1}a+ab_{3}cad_{3}a)^{*}}$.
It can be seen that $L(R)$ is controllable with respect to $L(G)$
and $\Sigma_{uc}$ and $L(R)$ is $C\&P$ co-observable with respect
to $\Sigma_{ci}$ and $\Sigma_{oi}$, where $i=1, 2$. In addition,
we can obtain $det(R)$ (Fig. \ref{ccls} (Left)), which implies
there is a bisimulation $\phi$ such that $G||det(R) \cong_{\phi}
R$. According to Theorem \ref{conjunctive}, there exist
decentralized bisimilarity supervisors for the conjunctive
architecture.

Then, $\mathcal{S}_{1}$ and $\mathcal{S}_{2}$ can be constructed
as below. The automata $S_1$ and $S_2$ can be found in Fig.
\ref{clsup}. Further, the local decision maps $\psi_{1}$ and
$\psi_{2}$ are described as follows.


\[
  \psi_{1}(y) = \left\{ {\begin{array}{*{20}c}
   \{a, c, b_3\} & {y=0,2,3,4,5,8,9, z_{d1}};  \\
   \{a, c, b_1, b_2, b_3\} & {y=1} ;\\
   \{a, c, b_3, d_1, d_3\} & {y=6} ;\\
   \{a, c, b_3, d_2\} & {y=7}.
\end{array}} \right.
\]

\[
  \psi_{2}(y) = \left\{ {\begin{array}{*{20}c}
   \{a, c, b_1, b_2, d_1, d_2\} & {y=0', 2', 3', 5', z_{d2}};  \\
   \{a, c, b_1, b_2, b_3, d_1, d_2\} & {y=1'};\\
    \{a, c, b_1, b_2, d_1, d_2, d_3\} & {y=4'};\\
\end{array}} \right.
\]


Then, the supervised system is shown in Fig. \ref{ccls} (Right).
It can be verified that $cl(\mathcal{S}_1,
\mathcal{S}_2)/_{\psi_{fc}}G\cong_{\phi_1\cup \phi_1^{-1}} R$,
where $\phi_1=\{((x_0, 0, 0'), q_0), ((x_1, 1,1'), q_1), ((x_2, 1,
1'), q_2),$\\$ ((x_3, 2,1'), q_3), ((x_3, 2,1'), q_5),((x_3, 3,
1'), q_4), ((x_3, 3, 1'), q_6), ((x_4, 1, 3'), q_3), ((x_4,$\\$ 1,
3'), q_5), ((x_4, 3, 1'), q_4), ((x_4, 3, 1'), q_6), ((x_5, 4,
2'), q_7), ((x_5, 5, 2'),q_8)((x_6, 6, 4'),$\\$q_9), ((x_6, 7,
4'), q_{11}),((x_6, 7, 4'), q_{12}), ((x_7, 6, 4'), q_{10}),
((x_7, 7, 4'), q_{11}), ((x_7, 7, 4'), q_{12}),$\\$ ((x_8, 8, 4'),
q_{13}), ((x_9, 6, 5'), q_{13}), (( x_8, 9, 4'), q_{13}), ((x_9,
9, 4'), q_{13}), ((x_8, 8, 4'), q_{14}), $\\$((x_9, 6, 5'),
q_{14}), ((x_8, 9, 4'), q_{14}), ((x_9, 9, 4'), q_{14})\}$ and
$\psi_{fc}$ is defined as (\ref{conj global dec}).


If we consider $det(R)$ (Fig. \ref{ccls} (Left)) as the
specification, it can be seen that $G||det(R)$ is not bisimilar to
$R$. Therefore, we can not find a solution for the decentralized
bisimilarity control problem. However, we can achieve the language
equivalence for the decentralized control problem since $L(R)$ is
$C\&P$ co-observable with respect to $L(G)$, $\Sigma_{oi}$ and
$\Sigma_{ci}$ for $i=1, 2$. Hence, the decentralized control for
language equivalence is easier the decentralized control for
bisimulation equivalence.

\begin{figure}[!htb]
\begin{center}
\includegraphics*[scale=.70]{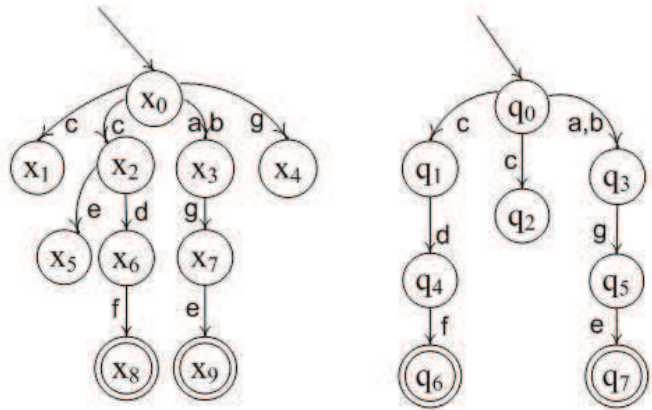}
\caption{Plant $G$ (Left) and Specification $R$ (Right) of Example
2} \label{dps}
\end{center}
\end{figure}

\begin{example}
Consider a plant $G$ and a specification $R$, which are shown in
Fig. \ref{dps}. Assume $i=1, 2$, $\Sigma_{o1}=\{a,c, d, e, f\}$,
$\Sigma_{o2}=\{b, c, d, e, f\}$, $\Sigma_{c1}=\{c, e, f, g\}$ and
$\Sigma_{c2}=\{d, e, f, g\}$. We could like to design
decentralized supervisors $\mathcal{S}_1$ and $\mathcal{S}_2$ with
a global decision rule $\psi_{f}$ such that $cl(\mathcal{S}_1,
\mathcal{S}_2)/_{\psi_{f}}G \cong R$.
\end{example}

\begin{figure}[!htb]
\begin{center}
\includegraphics*[scale=.70]{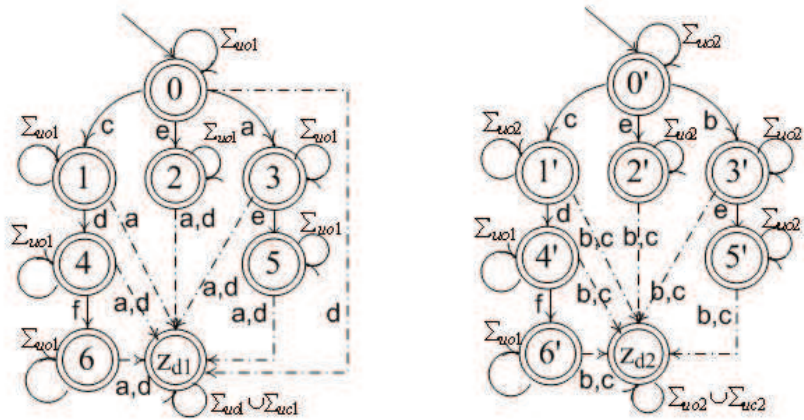}
\caption{The Automata of $\mathcal{S}_{1}$ (Left) and
$\mathcal{S}_{2}$ (Right) of Example 2} \label{dlsup}
\end{center}
\end{figure}

For $g \in \Sigma_{c1} \cap \Sigma_{c2}$, we have  $g \in
L(G)-L(R)$, $b \in P_{1}^{-1}P_{1}(\epsilon)$ and $a \in
P_{2}^{-1}P_{2}(\epsilon)$ such that $ag, bg \in L(R)$, where
$i=1, 2$. Therefore, $L(R)$ is not $C\&P$ co-observable with
respect to $L(G)$, $\Sigma_{oi}$ and $\Sigma_{ci}$, where $i=1,
2$. Thus, there does not exist a set of decentralized bisimilarity
supervisors for the disjunctive architecture. However, $L(R)$ is
$D\&A$ co-observable with respect to $L(G)$, $\Sigma_{oi}$ and
$\Sigma_{ci}$, where $i=1, 2$. Moreover, $L(R)$ is language
controllable with respect to $L(G)$ and $\Sigma_{uc}$ and
$G||det(R) \cong R$. It implies the existence of decentralized
supervisors for the disjunctive architecture to achieve the
bisimulation equivalence between the supervised system and the
specification.

Decentralized bisimilarity supervisors $\mathcal{S}_{1}=(S_1,
\psi_1)$ and $\mathcal{S}_{2}=(S_2, \psi_2)$ are designed
according to (\ref{con sup}) and (\ref{dis sup dec}), in which
$S_1$ and $S_2$ are shown in Fig. \ref{dlsup} and $\psi_1$ and
$\psi_2$ are presented as below.

\[
  \psi_{1}(y) = \left\{ {\begin{array}{*{20}c}
   \{a, b, c, e\} & {y=0};  \\
   \{a, b, e, g\} & {y=3} ;\\
   \{a, b, d\} & {y=1} ;\\
   \{a, b, f\} & {y=4};\\
    \{a, b\} & {y=2, 5, 6, z_{d1}}.\\
\end{array}} \right.
\]

\[
  \psi_{2}(y) = \left\{ {\begin{array}{*{20}c}
   \{a, b, c, e\} & {y=0'};  \\
   \{a, b, g, e\} & {y=3'};\\
   \{a, b, d\} & {y=1'};\\
   \{a, b, f\} & {y=4'};\\
   \{a, b, c\} & {y=2',5',6', z_{d2}}.\\
\end{array}} \right.
\]

\begin{figure}[!htb]
\begin{center}
\includegraphics*[scale=.70]{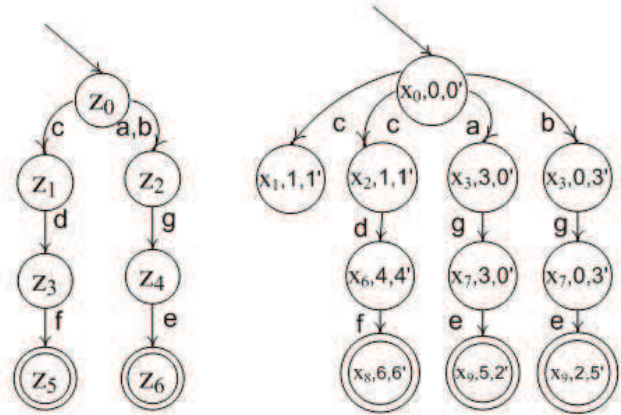}
\caption{$det(R)$ (Left) and Supervised System
$cl(\mathcal{S}_{1}, \mathcal{S}_2)/_{\psi_{fd}}G$ of Example 2}
\label{dcls}
\end{center}
\end{figure}

With $\mathcal{S}_{1}$ and $\mathcal{S}_{2}$, we obtain the
supervised system $cl(\mathcal{S}_1, \mathcal{S}_2)/_{\psi_{fd}}G$
(Fig. \ref{dcls} (Right)), where $\psi_{fd}$ is defined as
(\ref{disj global dec}). Let $\phi_1=\{((x_0, 0, 0'), q_0), ((x_2,
1, 1'), q_1), ((x_1, 1,$\\$ 1'), q_2), ((x_3, 3, 0'), q_3),
((x_3,0, 3'), q_3),  ((x_6,4, 4'), q_4),  ((x_7, 3,0'), q_5),
((x_7, 0, 3'), q_5), $\\$((x_8, 6, 6'),  q_6), ((x_8, 6, 6'),
q_7), ((x_9, 5, 2'), q_6), ((x_9, 5, 2'), q_7), ((x_9, 2, 5'),
q_6), ((x_9, 2, 5'), q_7)\}$. Therefore, $cl(\mathcal{S}_1,
\mathcal{S}_2)/_{\psi_{fd}}G \cong_{\phi_1\cup \phi_1^{-1}} R$.

\begin{figure}[!htb]
\begin{center}
\includegraphics*[scale=.70]{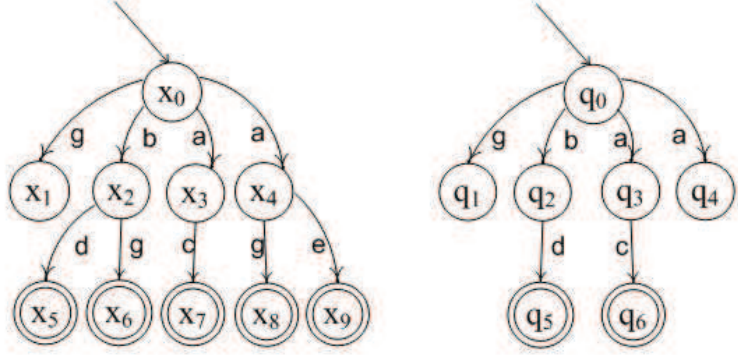}
\caption{Plant $G$ (Left) and Specification $R$ (Right) of Example
3} \label{cnotdps}
\end{center}
\end{figure}

\begin{example}
Consider a plant $G$ and a specification $R$, which are shown in
Fig. \ref{cnotdps}. Let $i=1, 2$, $\Sigma_{o1}=\{a,c\}$,
$\Sigma_{o2}=\{b, d\}$, $\Sigma_{c1}=\{g, e\}$ and
$\Sigma_{c2}=\{g, c, d\}$. The aim of control is to design
decentralized supervisors $\mathcal{S}_1$ and $\mathcal{S}_2$ with
a global decision fusion rule $\psi_{f}$ such that
$cl(\mathcal{S}_1, \mathcal{S}_2)/_{\psi_{f}}G \cong R$.
\end{example}

\begin{figure}[!htb]
\begin{center}
\includegraphics*[scale=.70]{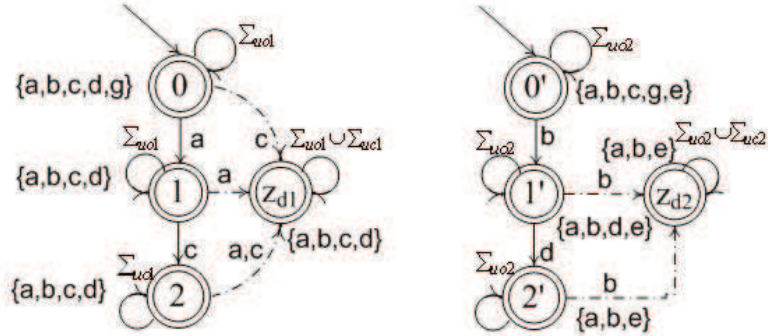}
\caption{The Automata of $\mathcal{S}_{1}$ (Left) and
$\mathcal{S}_{2}$ of Example 3} \label{cnotdlsup}
\end{center}
\end{figure}

For $g \in \Sigma_{c1} \cap \Sigma_{c2}$, we have $g \in L(R)$.
However, there exist $bg \in [(P_{1}^{-1}P_{1}(\epsilon)\cap
L(R))g \cap L(G)]$ and $ag \in [(P_{2}^{-1}P_{2}(\epsilon)g \cap
L(R))\sigma \cap L(G)]$ such that $ag, bg \notin L(R)$. Therefore,
$L(R)$ is not $D\&A$ co-observable with respect to $L(G)$,
$\Sigma_{oi}$ and $\Sigma_{ci}$, where $i=1, 2$, which implies
there does not exist a set of decentralized bisimilarity
supervisors for the disjunctive architecture.

\begin{figure}[!htb]
\begin{center}
\includegraphics*[scale=.70]{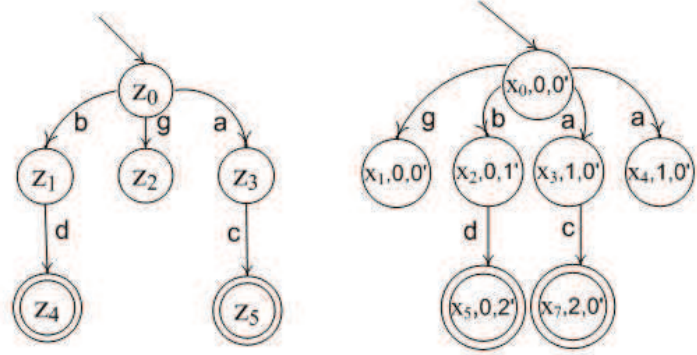}
\caption{$det(R)$ (Left) and Supervised System
$cl(\mathcal{S}_{1}, \mathcal{S}_2)/_{\psi_{fc}}G$ of Example 3}
\label{cnotdcls}
\end{center}
\end{figure}

However, $L(R)$ is $C\&P$ co-observable with respect to $L(G)$,
$\Sigma_{oi}$ and $\Sigma_{ci}$. In addition, $G||det(R) \cong R$
and $L(R)$ is language controllable with respect to $L(G)$ and
$\Sigma_{uc}$ and $det(R)$ is in Fig. \ref{cnotdcls} (Left).
Therefore, there exist decentralized bisimilarity supervisors for
the conjunctive architecture.

Refer to (\ref{con sup}) and (\ref{con sup dec}), we can design
decentralized supervisors $\mathcal{S}_{1}=(S_1, \psi_1)$ and
$\mathcal{S}_{2}=(S_2, \psi_2)$ as shown in Fig. \ref{cnotdlsup}.
The supervised system $cl(\mathcal{S}_1,
\mathcal{S}_2)/_{\psi_{fc}}G$ can be seen in Fig. \ref{cnotdcls},
where $\psi_{fc}$ is defined as (\ref{conj global dec}).
Therefore, $cl(\mathcal{S}_1, \mathcal{S}_2)/_{\psi_{fc}}G
\cong_{\phi_1 \cup \phi_1^{-1}} R$, where $\phi_1=\{((x_0, 0, 0'),
q_0), ((x_1, 0, 0'), q_1),((x_2, 0, 1'), q_2),  ((x_3, 1, 0'),
q_3), ((x_4, 1,$\\$ 0'), q_4), ((x_5, 0, 2'), q_5), ((x_7, 2, 0'),
q_5), ((x_5, 0, 2'), q_6), ((x_7, 2, 0')), q_6),  ((x_1, 0,
0'),q_4),$\\$ ((x_4, 1, 0'), q_1)\}$.

\begin{figure}[!htb]
\begin{center}
\includegraphics*[scale=.70]{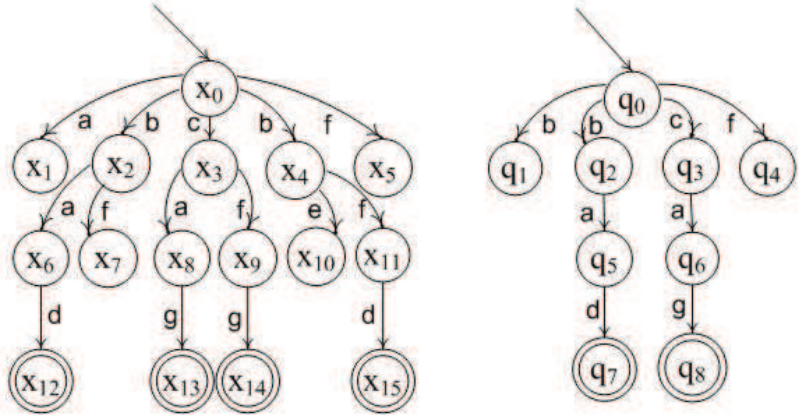}
\caption{Plant $G$ (Left) and Specification $R$ (Right) of Example
4} \label{cdps}
\end{center}
\end{figure}

\begin{example}
Consider a plant $G$ and a specification $R$, which are shown in
Fig. \ref{cdps}. Let $i=1, 2$, $\Sigma_{o1}=\{b, e\}$,
$\Sigma_{o2}=\{c, d\}$, $\Sigma_{c1}=\{a, f, e\}$ and
$\Sigma_{c2}=\{a, f\}$. In the following, we investigate the
problem whether there exist decentralized supervisors
$\mathcal{S}_1$ and $\mathcal{S}_2$ with a global decision fusion
rule $\psi_{f}$ such that $cl(\mathcal{S}_1,
\mathcal{S}_2)/_{\psi_{f}}G \cong R$ or not.
\end{example}

\begin{figure}[!htb]
\begin{center}
\includegraphics*[scale=.70]{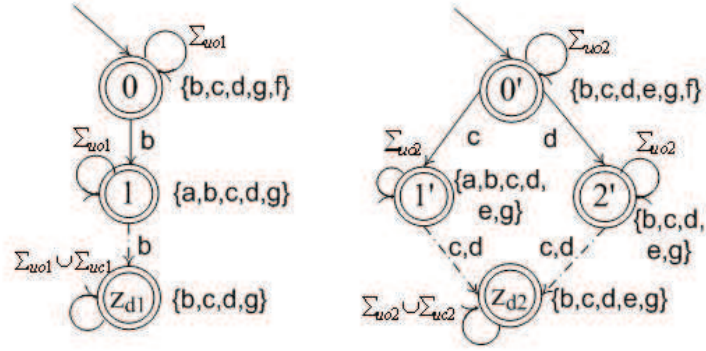}
\caption{Decentralized Supervisors $\mathcal{S}_{1}$ (Left) and
$\mathcal{S}_{2}$ (Right) of Example 4} \label{cdlsup}
\end{center}
\end{figure}

For $a \in \Sigma_{c1} \cap \Sigma_{c2}$, we have $a \notin L(R)$.
However, there exist $c \in P_{1}^{-1}P_{1}(\epsilon)$ and $b \in
P_{2}^{-1}P_{2}(\epsilon)$ such that $ca, ba \in L(R)$. Thus,
$L(R)$ is not $C \&P$ co-observable with respect to $L(G)$,
$\Sigma_{oi}$ and $\Sigma_{ci}$, where $i=1, 2$.

On the other side, we have $f \in L(R)$, $cf \in
[(P_{1}^{-1}P_{1}(\epsilon)\cap L(R))f \cap L(G)]\setminus L(R)$
and $bf \in [(P_{2}^{-1}P_{2}(\epsilon)\cap L(R))f \cap
L(G)]\setminus L(R)$ for $f \in \Sigma_{c1} \cap \Sigma_{c2}$.
Therefore, $L(R)$ is not $D\&A$ co-observable with respect to
$L(G)$, $\Sigma_{oi}$ and $\Sigma_{ci}$, where $i=1, 2$. By
Theorem \ref{conjunctive} and Theorem \ref{disjunctive}, there
does not exist a set of decentralized bisimilarity supervisors for
both the conjunctive architecture and the disjunctive
architecture.

Let $\Sigma=\Sigma_{ce} \cup \Sigma_{cd}$, where $\Sigma_{ce}=\{f,
e\}$ and $\Sigma_{cd}=\{a\}$. Then, $\Sigma_{ce1}=\Sigma_{ce} \cap
\Sigma_{c1}=\{f, e\}$, $\Sigma_{ce2}=\Sigma_{ce} \cap
\Sigma_{c2}=\{f\}$, $\Sigma_{cd1}=\Sigma_{cd} \cap
\Sigma_{c1}=\{a\}$ and $\Sigma_{cd2}=\Sigma_{cd} \cap
\Sigma_{c2}=\{a\}$. It can be easily verified that $L(R)$ is
co-observable with respect to $L(G)$, $\Sigma_{oi}$,
$\Sigma_{cei}$ and $\Sigma_{cdi}$, where $i=1, 2$. In addition,
$L(R)$ is language controllable with respect to $L(G)$ and
$\Sigma_{uc}$ and $G||det(R) \cong R$. Therefore, we can find
decentralized bisimilarity supervisors for the general
architecture.

\begin{figure}[!htb]
\begin{center}
\includegraphics*[scale=.70]{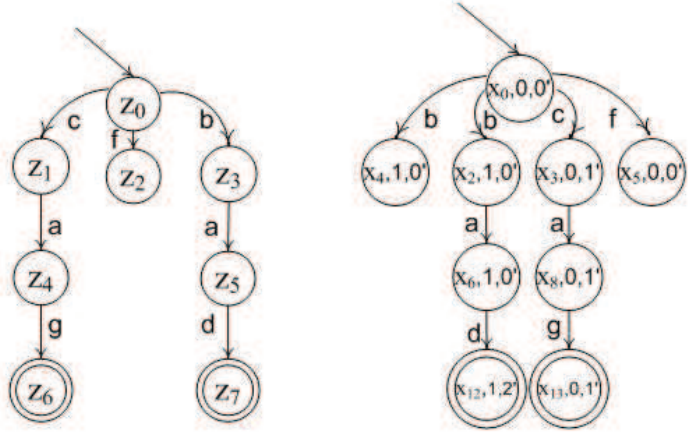}
\caption{$det(R)$ (Left) and Supervised System
$cl(\mathcal{S}_{1}, \mathcal{S}_2)/_{\psi_{fg}}G$ of Example 4}
\label{cdcls}
\end{center}
\end{figure}

Next, decentralized supervisors $\mathcal{S}_{1}=(S_1, \psi_1)$
and $\mathcal{S}_{2}=(S_2, \psi_2)$ are designed by using
(\ref{con sup}) and (\ref{gen sup dec}) in Fig. \ref{cdlsup}. It
can be seen that $cl(\mathcal{S}_1, \mathcal{S}_2)/_{\psi_{fg}}G
\cong_{\phi_1 \cup \phi_1^{-1}} R$ (Fig. \ref{cdcls}), where
$\phi_1=\{((x_0, 0, 0'), q_0), ((x_4, 1, 0'), q_1), ((x_2, 1, 0'),
q_2),  ((x_3, 0, 1'), q_3),$\\$ ((x_5, 0, 0'), q_4), ((x_6, 1,
0'), q_5), ((x_8, 0, 1'), q_6), ((x_{12}, 1, 2'), q_7), ((x_{13},
0, 1')), q_8), ((x_{12},$ \\$  1, 2'), q_8),((x_{13}, 0, 1')),
 q_7), ((x_4, 1, 0'), q_4), ((x_5, 0, 0'), q_1)\}$ and
 $\psi_{fg}$ is defined as (\ref{general global dec}).

\section{CONCLUSIONS}

The decentralized bisimilarity control of discrete event systems
was studied in this paper, where the plant and the specification
are modeled as nondeterministic automata and the supervisor is
modeled as a deterministic automaton. To formally capture
bisimulation equivalence, we propose an automata-based framework,
upon which a conjunctive architecture, a disjunctive architecture
and a general architecture were constructed for decentralized
bisimilarity control with respect to different decision fusion
rules. Then, necessary and sufficient conditions for the existence
of a set of $\Sigma_{uci}-compatible$ and
$\Sigma_{uoi}-compatible$ bisimilarity supervisors were presented
respectively under above three architectures. It was shown that
these conditions can be verified with exponential complexity.
Furthermore, when the existence condition holds, we provided a
synthesis method to design the decentralized bisimilarity
supervisors.

With the results of this paper, we can further investigate the
synthesis of supermal/infimal subspecifications when the existence
conditions are not satisfied. In addition, we can also study the
decentralized bisimilarity control problem by allowing
nondeterministic supervisors under the proposed framework. These
problems will be considered in our subsequent work.

%

\bibliographystyle{model5-names}
\bibliography{automatica2009,Ref_10}
\end{document}